\documentclass[iop,apj]{emulateapj}

\usepackage{amssymb,subfigure,rotating}

\shortauthors{Tombesi et al.}

\begin{document}


\title{Evidence for ultra-fast outflows in radio-quiet AGNs: II - detailed photo-ionization modeling of Fe K-shell absorption lines}


\author{F. Tombesi$^{1,2,3,4}$, M. Cappi$^{4}$, J.~N. Reeves$^{5}$, G.~G.~C. Palumbo$^{3}$, V. Braito$^{6}$, and M. Dadina$^{4}$}
\affil{$^1$ X-ray Astrophysics Laboratory, NASA/Goddard Space Flight Center, Greenbelt, MD 20771, USA}
\affil{$^2$ Department of Astronomy and CRESST, University of Maryland, College Park, MD 20742, USA; ftombesi@astro.umd.edu}
\affil{$^3$ Dipartimento di Astronomia, Universit\`a di Bologna, Via Ranzani 1, I-40127 Bologna, Italy}
\affil{$^4$ INAF-IASF Bologna, Via Gobetti 101, I-40129 Bologna, Italy}
\affil{$^5$ Astrophysics Group, School of Physical and Geographical Sciences, Keele University, Keele, Staffordshire ST5 5BG, UK}
\affil{$^6$ Department of Physics and Astronomy, University of Leicester, University Road, Leicester LE1 7RH, UK}




\begin{abstract}

X-ray absorption line spectroscopy has recently shown evidence for previously unknown Ultra-fast Outflows (UFOs) in radio-quiet AGNs. These have been detected essentially through blueshifted Fe~XXV/XXVI K-shell transitions. In the previous paper of this series we defined UFOs as those highly ionized absorbers with an outflow velocity higher than 10,000~km/s and assessed the statistical significance of the associated blueshifted absorption lines in a large sample of 42 local radio-quiet AGNs observed with \emph{XMM-Newton}. 
The present paper is an extension of that work.
First, we report a detailed curve of growth analysis of the main Fe~XXV/XXVI transitions in photo-ionized plasmas. Then, we estimate an average SED for the sample sources and directly model the Fe K absorbers in the \emph{XMM-Newton} spectra with the detailed \emph{Xstar} photo-ionization code. 
We confirm that the frequency of sources in the radio-quiet sample showing UFOs is $>$35\% and that the majority of the Fe K absorbers are indeed associated with UFOs. The outflow velocity distribution spans from $\sim$10,000~km/s ($\sim$0.03c) up to $\sim$100,000~km/s ($\sim$0.3c), with a peak and mean value of $\sim$42,000~km/s ($\sim$0.14c). The ionization parameter is very high and in the range log$\xi$$\sim$3--6~erg~s$^{-1}$~cm, with a mean value of log$\xi$$\sim$4.2~erg~s$^{-1}$~cm. The associated column densities are also large, in the range $N_H$$\sim$$10^{22}$--$10^{24}$~cm$^{-2}$, with a mean value of $N_H$$\sim$$10^{23}$~cm$^{-2}$.
We discuss and estimate how selection effects, such as those related to the limited instrumental sensitivity at energies above 7 keV, may hamper the detection of even higher velocities and higher ionization absorbers. We argue that, overall, these results point to the presence of extremely ionized and possibly almost Compton thick outflowing material in the innermost regions of AGNs.
This also suggests that UFOs may potentially play a significant role in the expected cosmological feedback from AGNs and their study can provide important clues on the connection between accretion disks, winds and jets.

\end{abstract}


\keywords{black hole physics --- galaxies: active --- galaxies: Seyfert --- line: identification --- plasmas --- X-rays: galaxies}



\section{Introduction}

Blueshifted Fe K absorption lines have been detected in recent years between 
7 and 10~keV in the X-ray spectra of several radio-quiet AGNs (e.g., Chartas 
et al.~2002; Chartas et al.~2003; Pounds
et al.~2003; Dadina et al.~2005; Markowitz et al.~2006; Braito et
al.~2007; Cappi et al.~2009; Reeves et al.~2009). These lines are
commonly interpreted as due to resonant absorption from Fe XXV--XXVI 
associated with a zone of circumnuclear gas photo-ionized by the
central X-ray source, with ionization parameter
log$\xi$$\sim$3--5~erg~s$^{-1}$~cm and column density
$N_H$$\sim$$10^{22}$--$10^{24}$~cm$^{-2}$.  The energies of these
absorption lines are systematically blueshifted and the corresponding
velocities can reach up to mildly relativistic values of
$\sim$0.2--0.4c.  
 These findings are important because they suggest the presence of previously unknown massive and highly ionized absorbing material outflowing from their nuclei, possibly connected with accretion disk winds/ejecta (e.g., King \& Pounds 2003; Proga \& Kallman 2004; Sim et al.~2008; Ohsuga et al.~2009; King 2010a; Sim et al.~2010) or the base of a possible weak jet (e.g., Ghisellini et al.~2004).

In particular, a uniform and systematic search for
blueshifted Fe K absorption lines in a large sample of 42 radio-quiet
AGNs observed with \emph{XMM-Newton} has been performed in Tombesi et
al.~(2010a, hereafter TA). This study allowed us to assess their global 
detection significance and to overcome any possible publication bias. For clarity, there we defined Ultra-fast Outflows (UFOs) those highly ionized absorbers detected essentially through Fe XXV and Fe XXVI K-shell absorption lines with blueshifted velocities v$\geq$10,000~km/s ($\sim$0.03c). This was done to distinguish from the classical warm absorbers (WA) in AGNs, which are commonly detected in the soft X-ray band of more than half of radio-quiet objects. In fact, these absorbers are usually less ionized, are outflowing with velocities lower than $\sim$1000~km/s  and are supposed to have a different physical origin (e.g., Blustin et al.~2005; McKernan et al.~2007).

In TA we detected UFOs in $\ga$35\% of the sources, with outflow velocities even larger than 0.1c in $\sim$25\% of them, and observed also short time-scale variability of $\sim$days in a few cases. 
However, in TA, the narrow absorption lines were modeled with simple inverted Gaussians and the blueshifted velocities were estimated assuming a line identification essentially as Fe XXV/XXVI 1s--2p and 1s--3p transitions. Here we extend that work and model the Fe K absorbers using the detailed photo-ionization code \emph{Xstar} v.~2.2 (Kallman \& Bautista 2001), specifically developed for X-ray astronomy, which allows to derive a physically self-consistent estimate of the ionization parameter, column density and outflow velocity of the UFOs.  

In addition, Tombesi et al.~(2010b, hereafter TB) also detected, for the first time, UFOs in three out of five radio-loud AGNs observed by \emph{Suzaku}. These are in the form of a series of narrow absorption lines at energies greater than 7~keV. Their likely interpretation as blueshifted Fe XXV/XXVI K-shell resonance lines implies an origin from highly ionized gas outflowing with mildly relativistic velocities, in the range $v$$\simeq$0.04--0.15c. A fit with specific photo-ionization models give ionization parameters in the range log$\xi$$\simeq$4--5.6~erg~s$^{-1}$~cm and column densities of
$N_H$$\simeq$$10^{22}$--$10^{23}$~cm$^{-2}$. Their estimated location within
$\sim$0.01--0.3pc from the central super-massive black hole (SMBH) suggests a
likely origin related to accretion disk winds/ejecta and maybe the jet present in these sources. 

Depending on the absorber covering factor and geometry, the mass outflow rate 
of these 
UFOs can be comparable to the accretion rate and their kinetic power can
correspond to a significant fraction of the bolometric luminosity and
is comparable to the typical jet power of radio-loud sources.  
Therefore, they might
have the possibility to bring outward a significant amount of mass and
energy, which can have an important influence on the surrounding
environment (e.g., see review by Cappi 2006). In fact, feedback from
the AGN is expected to have a significant role in the evolution of the
host galaxy, such as the enrichment of the ISM or the reduction of star
formation (e.g., see review by Elvis 2006 and Fabian 2009). 
Moreover, the ejection of a
substantial amount of mass from the central regions of AGNs can also
inhibit the growth of the SMBHs and could also explain some fundamental relations (e.g., King 2010a, b, and references therein),
potentially affecting their evolution. 
Therefore, these UFOs can play a significant role in the expected 
cosmological feedback from AGNs and can give us further clues on the 
relation between the accretion disk and the formation of winds/jets in 
both radio-quiet and radio-loud sources.

This paper is structured as follows. In \S2 we describe a curve of growth analysis of the main Fe XXV/XXVI K-shell absorption lines. In \S3 we present the calculation of the average SED of the TA sample sources and the fits to the \emph{XMM-Newton} spectra with the photo-ionization code \emph{Xstar}. In \S4 we discuss the results of the fits and show the distribution of the ionization parameter, column density and outflow velocity of the UFOs. The conclusions are reported in \S5. In Appendix A we report the detailed photo-ionization modeling for each source and in Appendix B the $\chi^2$ statistic and absorber redshift contour plots for each observation. 
We refer to a successive paper III of this series (Tombesi et al. in prep.) for a detailed discussion of the physical properties, geometry and energetics of the UFOs here analyzed and also a comparison with the classical warm absorbers.

\section{Curve of growth analysis of the Fe K absorption lines}

Due to the intense X-ray luminosity of the central source, absorbers in the inner regions of AGNs are mainly photo-ionized by this radiation instead of being locally ionized by thermal collisions (e.g., Kinkhabwala et al.~2002; Crenshaw et al.~2003).
In this way, they can reach very high levels of ionization maintaining relatively low and constant temperatures of $T$$\sim$$10^5$--$10^6$~K (e.g., Nicastro et al.~1999; Bianchi et al.~2005). 
In order to properly parameterize the observed absorption features we need first to have a clear quantitative idea of the general dependence and effects of the different physical parameters of the photo-ionized plasma on the strength of the Fe K absorption lines. 
These effects and dependencies have already been studied by Bianchi et al.~(2005) and Risaliti et al.~(2005). Nevertheless, these authors were mainly interested in low velocity photo-ionized outflows in AGNs, instead of the ultra-fast outflows considered here, and used slightly different assumptions. However, the radiation processes are essentially the same in both cases and their results can be directly compared with ours. Therefore, in this section we describe a detailed curve of growth analysis of the highly ionized Fe K-shell absorption lines and then compare with what observed in the \emph{XMM-Newton} spectra of TA.

\subsection{Fe XXV and Fe XXVI K-shell lines}

We consider the most intense K-shell transitions of Fe XXV and Fe XXVI reported in Table~1. All line parameters throughout this paper have been taken from the NIST\footnote{http://physics.nist.gov/PhysRefData/ASD/index.html.} atomic database, if not otherwise stated. 

The most prominent Fe XXV lines are due to the He$\alpha$ (1s$^2$--1s2p) and He$\beta$ (1s$^2$--1s3p) transitions, which further subdivide into inter-combination (i) and resonance lines (r). At the moderate energy resolution of the \emph{XMM-Newton} European Photon Imaging Camera (EPIC) pn instrument (FWHM$\ga$150~eV at 6~keV) these fine structure line components are not distinguishable and they are blended. Therefore, we averaged their energies, weighted for the respective oscillator strengths. The consequent line energies are 6.697~keV for the Fe XXV He$\alpha$ and 7.880~keV for the Fe XXV He$\beta$, respectively, see Table~1.
The most intense lines from Fe XXVI are instead due to the Lyman series, in particular the Ly$\alpha$ (1s--2p) and Ly$\beta$ (1s--3p), each of which further subdivide into two resonance doublets, r$_1$ and r$_2$. Their average energies, weighted for their oscillator strengths, are 6.966~keV for the Fe XXVI Ly$\alpha$ and 8.250~keV for the Fe XXVI Ly$\beta$, respectively, see Table~1. 

It should be noted that these strong lines are always accompanied by further less intense, higher order K-shell resonances, such as: Fe XXV He$\gamma$ (1s$^2$--1s4p) at E$\simeq$8.295~keV, Fe XXV He$\delta$ (1s$^2$--1s5p) at E$\simeq$8.487~keV, Fe XXVI Ly$\gamma$ (1s--4p) at E$\simeq$8.701~keV and Fe XXVI Ly$\delta$ (1s--5p) at E$\simeq$8.909~keV. 
However, their detection is hampered by the fact that they are intrinsically weaker than the former, with oscillator strengths $\la$5\% and therefore with less intensity, and they are located at E$>$8~keV where the S/N and energy resolution of the EPIC pn spectra are worse. Therefore, here we do not present a curve of growth analysis of these weak lines but they are consistently taken into account when fitting the spectra using the \emph{Xstar} code (see \S3.2).

\subsection{Theoretical calculation}

We performed a curve of growth analysis of the most intense Fe XXV/XXVI K-shell absorption lines generated in a photo-ionized plasma. We used the line parameters reported in Table~1. This study allows us to derive a quantitative estimate of the dependence of the equivalent width (EW) of the lines, directly observable in the X-ray spectra, on the main physical parameters describing the absorbing material, such as the ionization parameter $\xi$, the column density $N_H$ and the velocity broadening parameter $b$. All calculations were performed separately for each line component listed in Table~1 and only their final EWs were summed.

The relative abundances of Fe XXV and Fe XXVI ions are calculated by means of the photo-ionization code \emph{Xstar} as a function of the ionization parameter and column density.
We assume a power-law ionizing continuum with $\Gamma$$=$2, which is the typical value for Seyfert galaxies in the X-rays (e.g., Bianchi et al.~2005; Dadina 2008 and references therein). 
We use the standard definition of the ionization parameter $\xi=L_{ion}/nr^{2}$, in units of erg~s$^{-1}$~cm (Tarter, Tucker \& Salpeter 1969), where $L_{ion}$ is the ionizing luminosity between 1~Ryd and 1000~Ryd (1~Ryd$=$13.6~eV), $n$ is the number density of the material and $r$ is the distance of the gas from the central source. This expression directly derives from the ionization balance equations and essentially represents the ratio between the density of the ionizing photons and the density of the absorbing material. 

We assume standard Solar abundances and in particular an iron abundance relative to hydrogen of  $A_{\mathrm{Fe}} = 3.16\times 10^{-5}$ (Asplund et al.~2009).
We assume a constant density of the gas, with the typical value used for the Broad Line Region of type 1 AGNs, $n$$=$$10^{10}$~cm$^{-3}$ (e.g., Crenshaw et al.~2003 and references therein). 
We did not fix the gas temperature but let the code calculate the thermal equilibrium. However, the resultant temperatures are always consistent with the typical value of $T$$\sim$$10^6$~K (e.g., Nicastro et al.~1999; Bianchi et al.~2005).
Finally, we limited the treatment to a Compton thin absorber, $N_H$$\leq$$10^{24}$~cm$^{-2}$, distributed on a geometrically thin shell, $\Delta r/r=N_H \xi^{1/2} (L_{ion}n)^{-1/2}<1$.

These assumptions are slightly different from those of Bianchi et al.~(2005) and Risaliti et al.~(2005). In particular, they derived the iron ions populations using a different photo-ionization code, CLOUDY (Ferland 2003), but considered the same power-law continuum with $\Gamma$$=$2 as in this work. They assumed a lower gas density of $n$$=$$10^6$~cm$^{-3}$ and fixed the gas temperature to $T$$=$$10^6$~K. They considered a slightly different iron abundance relative to hydrogen of $A_{\mathrm{Fe}} = 4.68\times 10^{-5}$ (Anders \& Grevesse 1989).
They also used a different definition of the ionization parameter, $U_x$, which however can be simply related to ours as log$\xi$$\simeq$4$+$log$U_x$, for the $\Gamma$$=$2 power-law continuum considered here.

To derive the EW of the absorption lines we followed an approach similar to that of Nicastro et al.~(1999) and Bianchi et al.~(2005). The EW of a resonant absorption line is defined as:

\begin{equation}
\mathrm{EW}=\int_{\, 0}^{\, +\infty} (1-e^{-\tau_{\nu}}) \mathrm{d}\nu \,,
\end{equation}
 
\begin{flushleft}
where $\tau_{\nu}$ is the dimensionless frequency specific optical depth of the considered transition:
\end{flushleft}  

\begin{equation}
\tau_{\nu}=\int_{\,0}^{\,L} \alpha_{\nu} \mathrm{d}s = n_l L \frac{\pi e^2}{m_e c} f_{lu} \Phi(\nu) \,.
\end{equation}

In the above equation $\alpha_\nu$ is the absorption coefficient at the frequency $\nu$, $n_l$ is the number density of ions of the given element that populate the lower level of the transition considered, $L$ is the linear extension of the material along the line of sight, $f_{lu}$ is the oscillator strength of the electron transition from the lower to the upper level and $\Phi(\nu)$ is the normalized Voigt profile of the line. We then directly integrated the Voigt profile, which can be written as (e.g., Rybicki \& Lightman 1979):

\begin{equation}
\Phi(\nu)= \frac{1}{\Delta\nu_{D} \sqrt{\pi}} \, H(a,u)\,,
\end{equation}

\begin{flushleft}
where $H(a,u)$ is the Voigt function:
\end{flushleft}

\begin{equation}
H(a,u)= \frac{a}{\pi} \int_{\, -\infty}^{\, +\infty} \frac{e^{-y^{2}}}{a^2 + (u-y)^{2}} \mathrm{d}y \,.
\end{equation}

in which $a=\gamma/(4\pi\Delta\nu_{D})$ and $u=(\nu -\nu_0)/\Delta\nu_{D}$. 

This is essentially a convolution of the Gaussian and Lorentzian line profiles. The Gaussian profile, which represents the core of the line, depends on the Doppler broadening $\Delta\nu_{D}=\nu_0 b/c$, where $\nu_0$ is the rest-frame specific frequency of the transition and $b$ is the velocity broadening parameter. 
The latter is the square root of the quadratic sum of the thermal velocity of the ions, $\mathrm{v}_{\mathrm{th}}^2=2kT/m$, where $T$ is the gas temperature and $m$ the ion mass, and the gas turbulent velocity, or velocity dispersion along the line of sight, v$_{\mathrm{tu}}$: $b = \left(\mathrm{v}_{\mathrm{th}}^2 + \mathrm{v}_{\mathrm{tu}}^{2}\right)^{1/2}$.

The expected temperature of a highly ionized gas in photo-ionization equilibrium in the central regions of AGNs is of the order of $T$$\sim$$10^6$~K (e.g., Bianchi et al.~2005 and references therein) and the opacity is dominated by Fe ions. We can estimate that the consequent thermal velocity of the gas is less than 20~km/s and therefore its contribution to the velocity broadening is negligible, less than 2\%, for a typical value of the turbulent velocity of $\sim$1000~km/s (e.g., Nicastro et al.~1999; Bianchi et al.~2005). Therefore, the velocity broadening parameter is dominated by the turbulence component and this fact is also strengthened by the linear dependence $b\propto$~v$_{\mathrm{tu}}$, instead of $b\propto T^{1/2}$.
On the other hand, the Lorentzian profile contributes to the broadening of the line wings and is a function of the natural broadening parameter $\gamma$, which in turn depends on the Einstein coefficient for spontaneous emission for the transition considered, $A_{ul}$ (e.g., Rybicki \& Lightman 1979).

The curve of growth of the Fe XXV He$\alpha$ and Fe XXVI Ly$\alpha$ lines for a fixed ionization parameter log$\xi$$=$3.5~erg~s$^{-1}$~cm are reported in the upper and lower panels of Fig.~1, respectively. 
For relatively low column densities, less than $\sim$$10^{22}$~cm$^{-2}$, the curves follow a linear relation and the increase in EW is proportional to $N_H$, EW $\propto\tau\propto N_H f_{lu}$. For larger columns, instead, saturation effects become more intense and the increase in EW is lower. 
However, here we can clearly see the effects of increasing the velocity broadening, from 20~km/s up to 10,000~km/s. The Gaussian line core broadens and the column density needed to saturate the line is larger. Therefore, the linear branch of the curve of growth spans over a larger range of $N_H$ before saturating and the lines can reach higher EWs. In fact, we can see that for column densities of $N_H$$\sim$$10^{23}$--$10^{24}$~cm$^{-2}$ the EW of both lines can reach values of $\sim$10~eV for $b$$\simeq$20~km/s and up to a few $\sim$100~eV for $b$$\simeq$10,000~km/s.

As already noted before, given that the temperature of the photo-ionized gas remains almost constant at the value of $T$$\sim$$10^6$~K (e.g., Nicastro et al.~1999; Bianchi et al.~2005), which corresponds to a thermal velocity of the Fe ions lower than $\sim$20~km/s, the broadening is essentially dominated by an increase in the turbulence component. 
We note that absorption lines with turbulent velocities as high as $\sim$1000~km/s are shown by many broad-line AGNs (e.g., Crenshaw et al.~2003 and references therein) and such values are easily reached in gas moving close to SMBHs (e.g. Nicastro et al.~1999; Bianchi et al.~2005 and references therein).

The dependence of the EW of the Fe XXV He$\alpha$ and Fe XXVI Ly$\alpha$ absorption lines with respect to the ionization parameter is shown in Fig.~2 for a set of given column densities and velocity broadening.  
The location of the peak of the distributions depends on the specific ionic population for different ionization states. In fact, the highest abundance for Fe XXV is reached around log$\xi$$\simeq$3~erg~s$^{-1}$~cm. Consequently, Fe XXVI peaks at higher ionization levels of log$\xi$$\simeq$3.5--4~erg~s$^{-1}$~cm. 
Moreover, it should be noted that the EW distribution of Fe XXV He$\alpha$ is narrow, with a sharp rise and decline between log$\xi$$\sim$2.5--4~erg~s$^{-1}$~cm. Instead, that of Fe XXVI Ly$\alpha$ is much broader, with a sharp rise at log$\xi$$\simeq$3~erg~s$^{-1}$~cm and a tail extending even up to log$\xi$$\simeq$6~erg~s$^{-1}$~cm.
Therefore, the Fe XXVI transitions clearly dominate for ionization levels higher than log$\xi$$\simeq$4~erg~s$^{-1}$~cm.
It is also important to note the rise of the EW following an increase of the velocity parameter, which is equivalent to say for higher turbulent velocities. In fact, increasing the velocity broadening $b$ from 100~km/s up to 5000~km/s, the resultant value at the peak can reach values from a few $\sim$10~eV up to a few $\sim$100~eV for $N_H$$\sim$$10^{23}$--$10^{24}$~cm$^{-2}$ (see Fig.~2). This, again, shows that the increase in turbulent velocity is an efficient mechanism to reduce the line saturation.

Fig.~3 shows the ratio between the EWs of the Fe XXV He$\alpha$ and Fe XXVI Ly$\alpha$ absorption lines as a function of the ionization parameter and for different column densities.
We can note that for an ionization level of log$\xi$$\sim$3.5~erg~s$^{-1}$~cm the ratio is $\sim$1 and therefore we expect to observe both lines in the spectrum with comparable EWs. 
Instead, for lower or higher values of the ionization parameter the Fe XXV He$\alpha$ or the Fe XXVI Ly$\alpha$ respectively dominate. In these cases we expect to observe essentially only one of these two strong lines.
However, we can note that increasing the column density the ratio tends to approach unity for a wider range of ionization levels. In the limit of $N_H$$\sim$$10^{24}$~cm$^{-2}$, there is a plateau in the range log$\xi$$\sim$3--4~erg~s$^{-1}$~cm in which the ratio is $\sim$1. Therefore, also in this case we would expect to observe both lines in the X-ray spectrum with similar EWs.  
This is again an effect due to the line saturation and it is less intense for higher velocity parameters.  

The EW ratios of Fe XXV He$\alpha$/He$\beta$ and Fe XXVI Ly$\alpha$/Ly$\beta$ for different column densities and velocity parameters are shown in the upper and lower panels of Fig.~4, respectively.  The ionization level is fixed to log$\xi$$=$3.5~erg~s$^{-1}$~cm for reference. We can note that for column densities lower than $\sim$$10^{21}$--$10^{22}$~cm$^{-2}$ the ratios are essentially dictated by the ratios of the different oscillator strengths of the transitions, $\sim$5 (see Table~1). However, there is a tendency to approach unity for higher column densities.
This effect is again due to the line saturation. In fact, for a fixed velocity broadening, the transitions with higher oscillator strengths, in this case He$\alpha$ and Ly$\alpha$, quickly reach high EWs and therefore their cores saturate earlier than for the weaker lines, whose EWs can instead still grow. 
It follows that for high column densities, $N_H$$\ga$$10^{22}$~cm$^{-2}$, there is the possibility to observe both He$\alpha$/He$\beta$ and Ly$\alpha$/Ly$\beta$ absorption lines with comparable EWs in the spectrum.
This effect is however mitigated by an increase in the velocity broadening. As clearly shown in the upper and lower panels of Fig.~4, higher values of $b$ cause the decrease trend of the ratios to shift toward higher columns and the initial values are maintained for a wider range of $N_H$.  

Finally, we investigate the effects of changing different input parameters to the \emph{Xstar} code on the curve of growth of the Fe K absorption lines.
First, we checked the effects of assuming different continuum slopes.
We find that changes of the power-law photon index in the wide range of values $\Gamma$$=$1.5--2.5 provide EWs which are always consistent within 10\%. In particular, this essentially causes a slight shift of the peak of the EW distribution towards higher/lower ionization parameters for flat/steep slopes.
This effect is mainly due to the fact that, given the same bolometric luminosity, in the flat/steep power-law cases we have more/less high energy photons at E$>$6~keV that can effectively ionize iron atoms and produce Fe XXV/XXVI absorption lines. Therefore, to obtain the same iron ionic populations, steep spectra require slightly higher ionization levels than flat ones.  
From the definition of the ionization parameter, and assuming constant luminosity and density, this can be translated into the need for the material to be closer to the X-ray radiation source for steep spectra in order to get more hard X-ray flux to increase the Fe XXV/XXVI populations.   

We also tested that the inclusion of a possible UV-bump between $\sim$1--100~eV in the SED yields negligible effects to the Fe K band based results. In fact, as already stated by McKernan et al.~(2003), the presence or absence of the bump in the SED does not change significantly the parameters of the photo-ionized gas in the Fe K band because the main driver here is again the ionizing continuum in the X-rays.
We checked that the gas density does not play an important role in the overall modeling and the results are basically the same for a wide range of values, $n$$\sim$$10^2$--$10^{12}$~cm$^{-3}$ (e.g., Yaqoob et al.~2003; Bianchi et al.~2005). This is partially due to the fact that changing the density is equivalent to change the distance or luminosity, while leaving the same ionization parameter.

Finally, it is important to note that, even using slightly different assumptions, our results showed in Fig.~1 to Fig.~4 are quantitatively consistent overall with those showed in Fig.~1 to Fig.~6 of Bianchi et al.~(2005) and Fig.~4 of Risaliti et al.~(2005). 
The consistency of these independent studies means that the results do not drastically depend on the different assumptions and photo-ionization codes employed and this puts our final conclusions on a solid theoretical ground.

\section{Detailed photo-ionization modeling of the Fe K absorption lines}

In this section we describe the direct fit of the blueshifted Fe K absorption features detected by us in TA at E$>$6.4~keV in the spectra of the \emph{XMM-Newton} EPIC pn with the detailed photo-ionization code \emph{Xstar}. This allows us to derive estimates of the fundamental parameters of the absorbing material, such as the ionization level, column density and outflow velocity, in a physically self-consistent way.

\subsection{The Spectral Energy Distribution}

A good knowledge of the Spectral Energy Distribution (SED) of each source is important for the successive photo-ionization modeling of the Fe K absorbers and is one of the main input quantities required by the \emph{Xstar} code. The shape of the SED in different energy bands depends on the emission processes which dominate the production of that radiation.
As stated by Elvis et al.~(1994), a single, nearly horizontal power-law is indeed a good phenomenological representation of the SED of typical AGNs, from the IR up to the X-rays. However, some deviations can be observed. 

For instance, the \emph{millimeter break}, which is a drop in the power output in the sub-millimeter band compared to the optical. If the drop is $\sim$2 orders of magnitude the sources are classified as radio-loud, instead if higher, even $\sim$5--6 decades, they are labeled as radio-quiet. The latter are far more common, about $\sim$5--10 times more than the former (e.g., Kellerman et al.~1989 and references therein).
In particular, the radio-loudness parameter defined as $R_L=\textrm{log}(f_{5GHz}/f_{4400})>1$ is usually used to formally differentiate between radio-loud and radio-quiet AGNs, where $f_{5GHz}$ is the core 5 GHz radio flux and $f_{4400}$ is the flux at 4400 \AA, both in units of mJy (e.g., Kellerman et al.~1989 and references therein).   
This low energy emission is usually associated with the power-law synchrotron radiation from the relativistic radio jet. 

The \emph{UV bump} (or big blue bump) is a rise of the optical-UV continuum above the IR. It is often interpreted in terms of thermal emission from the putative accretion disk. In particular, the beginning of the bump is marked by an inflection between 1 and 1.5 $\mu$m, which is called the \emph{near-IR inflection point}. This is the only continuum feature which wavelength is well defined and it will be fundamental for the successive calculation of the average SED (e.g., Elvis et al.~1994). 

However, these "low energy" parts of the SED are almost negligible for our purposes of the photo-ionization modeling of highly ionized iron ions. Instead, it is more important to have a good knowledge of the SED in the X-ray band, especially at E$>$1~keV. 
The X-ray emission of radio-quiet AGNs is generally supposed to come from inverse Compton of optical-UV seed disk photons in a hot corona. This component is well represented by a power-law continuum with typical photon index $\Gamma$$\simeq$2, where $F_{E}$$\propto$$E^{-\Gamma}$~ph~s$^{-1}$ cm$^{-2}$~keV$^{-1}$, and exponential cut-off at E$\ga$100~keV (e.g., Haardt \& Maraschi 1991; Dadina 2008 and references therein).

As already stated, we need an estimate of the SED to calculate the \emph{Xstar} grid for the successive fitting of the Fe K absorption lines. 
However, given the large number of sources (42) and observations (101) of the TA sample, it would be too much time consuming to calculate one for each case. Moreover, the observed flux values at different energies have not been measured simultaneously and the SEDs should be corrected for possible intrinsic source variability. Thus, there is the need to find a reasonable approximation of the typical phenomenological SED.

We focus on the SEDs of the radio-quiet type 1 sources of our TA sample. 
This is due to the fact that possible obscuration and stellar contamination effects can prevent us from deriving a detailed knowledge of the intrinsic shape of the SED for the type 2s.
We will make the reasonable assumption that their central engine is analogous to that in type 1s and that the overall shape of their intrinsic SED is essentially the same. This is consistent with the widely accepted unification model for AGNs which states that the main difference between the two classes is essentially due to a different observing angle with respect to the line of sight (e.g., Antonucci 1993; Urry \& Padovani 1995 and references therein). 
The type 1s also constitute the majority of the objects of our sample, 35 over 42, and therefore they allow for better statistics.

We derive the average SED of the radio-quiet type 1 sources as follows: 1) we downloaded the observed SEDs of the 35 type 1 AGNs from the NED database\footnote{http://nedwww.ipac.caltech.edu}. These are composed by a collection of flux values measured from radio up to $\gamma$-rays. 2) we followed the approach of Elvis et al.~(1994) and normalized all the SEDs to the same arbitrary flux value at the near-IR inflection point at $\simeq$1.25 $\mu$m (E$\simeq$1~eV, see above).
Interestingly, the overall shape of the SEDs seems to be rather simple and is indeed similar for all the sources, see Fig.~5 (black lines).  
Then, 3), we calculated the mean flux value for each energy and derived an average SED for each object.
4) we averaged the flux values for each energy among all SEDs and derived a single average SED for all the sources.
Then, 5), we interpolated the fluxes between consecutive energy points with power-laws. In doing so, we noted that the slopes were consistent one with each other for several wide energy intervals and this suggested to us the identification of three main energy bands.
Finally, 6), we derived that essentially only four energy-flux points are required to properly describe the overall phenomenological SED and we calculated the power-law slopes among them. The values are reported in Table~2.

The final average phenomenological SED can be simply divided in three main intervals: from radio to millimetric ($\Gamma$$\sim$2), from millimetric to IR ($\Gamma$$\sim$0.1) and from IR to X-rays ($\Gamma$$\sim$2).  We remind the reader that the relation between the photon index ($\Gamma$) and spectral index ($\alpha$) is $\alpha=-\Gamma +1$.
As shown in Fig.~5 (red line), this indeed provides a good representation of the overall phenomenological SED of the type 1 AGNs.
It is important to note that the value of $\Gamma$$\sim$2 for the X-ray continuum, third interval in Table~2, is in agreement with the more detailed work of Elvis et al.~(1994) and with extensive broad-band X-ray spectral analyses (e.g., Dadina 2008 and references therein). Moreover, the mean phenomenological power-law slope in the E$=$4--10~keV band of the type 1 sources of TA is $<$$\Gamma$$>$$\sim$1.8. 
 This is also expected from the two phase Comptonization models, which describe the production mechanism of the X-ray radiation (e.g., Haardt \& Maraschi 1991), and $\Gamma$$\sim$2 is also usually assumed in photo-ionization studies of Fe K absorbers in AGNs (e.g., Bianchi et al.~2005; Cappi et al.~2009; TB and references therein).

\subsection{Fit of the \emph{XMM-Newton} EPIC pn spectra}

We consider our previous paper TA as the starting point for the modeling of the Fe K absorbers and we refer the reader to that for a detailed description of the data reduction, the continuum and emission lines modeling and the absorption lines detection significance. Here we consider again the E$=$3.5--10.5~keV energy band of the 28 \emph{XMM-Newton} EPIC pn observations in which narrow Fe K absorption features at E$>$6.4~keV were detected by TA, see their Table~A.2. The spectral analysis was carried out using the \emph{XMM-Newton} SAS~11.0 and \emph{heasoft} v.~6.10 packages and XSPEC v.~12.6.0. We binned the spectra to 1/5 of the FWHM of the EPIC pn using the SAS task \emph{specgroup} in order to sample the resolution of the detector. Given the X-ray brightness of the sources in the sample, this allows us to retain sufficient spectral information without the risks of oversampling the instrumental resolution. Given that FWHM$\simeq$160--180~eV in the Fe K band (Guainazzi et al.~2010) this is equivalent to an energy binning of $\sim$30--40~eV. However we checked that repeating the analysis considering a grouping of only 25 counts per bin and the use of the $\chi^2$ statistic, or alternatively not grouping the data at all and using the C-statistic (Cash 1979), provide equivalent results. The detailed description of the fits for the observations of each source is reported in Appendix A. The errors are reported at the 1$\sigma$ level, if not otherwise stated.

We started by fitting the data using the baseline models listed in Table~A.2 of TA with the absorption lines modeled as inverted Gaussians and derived an estimate of their velocity width, $\sigma_v$. This is important for the successive calculation of different \emph{Xstar} grids. As already stated in TA, the majority of the lines are unresolved and there we assumed widths of 10--100~eV. Given the limited spectral resolution of the \emph{XMM-Newton} EPIC pn in the considered Fe K band of FWHM$\simeq$160--180~eV (Guainazzi et al.~2010) and the S/N of the observations at E$>$7~keV, absorption lines with velocity widths lower than $\sim$2000--4000~km/s can not be resolved.
In fact, repeating the analysis, we could place only upper limits in the range $\sigma_v$$\la$3,000--10,000~km/s for the majority of them. However, in four cases we have been able to actually resolve the width of the features to $\sim$5,000~km/s, namely NGC~4151 (0402660201) $\sigma_v$$=$$5100^{+1800}_{-1400}$~km/s, NGC~4051 (0157560101) $\sigma_v$$=$$4700^{+1800}_{-1500}$~km/s, Mrk~766 (0304030301) $\sigma_v$$=$$4600^{+2700}_{-1800}$~km/s and PG~1211$+$143 (0112610101) $\sigma_v$$=$$4500^{+1500}_{-1200}$~km/s. These values are reported in Appendix A for each observation.   

The identification of spectral features in the Fe K band is relatively secure since only K-shell lines of heavy ions are expected at E$>$6~keV and those of iron are by far the most intense (e.g., Kallman et al.~2004). In TA we assumed a phenomenological identification with Fe~XXV He$\alpha$ or Fe XXVI Ly$\alpha$, favoring the latter if no contrary indications were present, and calculated the relative blueshifted velocity. There we explained the reasons supporting this choice, but postponed to this second paper for a conclusive test.

The possible contribution from lower ionization species of iron and the presence of ionized Fe K edges should be taken into account. These edges have rest-frame energies in the range from E$\simeq$7.1~keV to E$\simeq$9.3~keV, depending on the ionization state of iron, from neutral to H-like (e.g., Kallman et al.~2004). As a sanity check, in TA we tested that the alternative modeling of the absorption lines with simple sharp absorption edges (\emph{zedge} in XSPEC) did not significantly improve the spectral fits, as expected from the narrowness of the observed spectral features. Moreover, it is important to note that the commonly held view of sharp Fe K edges is an oversimplification of the real process and could lead to misleading results. In fact, it has been demonstrated that if the adequate treatment of the decay pathways of resonances converging to the K threshold is properly taken into account, the resulting edges are not sharp but smeared and broadened (Palmeri et al.~2002; Kallman et al.~2004). This effect can be negligible for neutral or extremely ionized iron (He/H-like) but is quite relevant for intermediate states, with energies in the range E$\simeq$7.2--9~keV. Furthermore, intense Fe K resonance absorption lines from different ionization states would be expected to accompany the edges. A proper physically self-consistent modeling of the absorbers in the Fe K band can only be performed using a detailed photo-ionization code. Here we use \emph{Xstar}.

For this reason, we derived three \emph{Xstar} grids considering column densities in the range $N_H$$=$$10^{20}$--$10^{24}$~cm$^{-2}$, ionization parameters in the interval log$\xi$$=$0--8~erg~s$^{-1}$~cm and turbulent velocities of 1000~km/s, 3000~km/s and 5000~km/s, respectively. We approximated the nuclear X-ray ionizing continuum with the average SED derived in \S3.1. In the input energy range of \emph{Xstar}, from E$=$0.1~eV to E$=$$10^6$~eV, this is essentially a power-law with $\Gamma$$=$2 and cut-off at E$>$100~keV, see Table~2. We assumed standard Solar abundances (Asplund et al.~2009). The free parameters for each \emph{Xstar} grid are the column density $N_H$, logarithm of the ionization parameter log$\xi$ and observed absorber redshift $z_o$. The observed absorber redshift is simply related to the intrinsic absorber redshift in the source rest frame $z_a$ as $(1+z_o)=(1+z_a)(1+z_c)$, where $z_c$ is the cosmological redshift of the source. The velocity can then be determined using the relativistic Doppler formula $1+z_a=(1-\beta/1+\beta)^{1/2}$, as $\beta=v/c$ is here required to be positive for an outflow.

Therefore, we fitted the data with the baseline models listed in Table~A.2 of TA, this time without the inclusion of the inverted Gaussian lines, and modeled the absorption features using \emph{Xstar}. For each observation, if the velocity width $\sigma_v$ of the lines was constrained, we performed a fit using the \emph{Xstar} grid with the closest turbulent velocity, which turned out to be possible only for the four cases reported above with $\sigma_v$$\simeq$5,000~km/s. Instead, if only an upper limit was derived, we performed fits using different \emph{Xstar} grids with turbulent velocities of 1000~km/s, 3000~km/s and 5000~km/s until these values were lower or equal to the line velocity width and then checked which one provides a better fit.

To find the \emph{Xstar} solution(s), i.e. the possible minima in the $\chi^2$ distribution, we stepped through the absorber redshift in small increments of $\Delta z$$=$$10^{-3}$ in the interval between $0.1$ to $-0.4$, leaving also the other parameters of the absorber, i.e. $N_H$ and log$\xi$, and the continuum free to vary and recorded the $\chi^2$ values. For the case of 1H0419-577 we stepped the redshift starting from $0.15$ because the source cosmological redshift is $z$$=$$0.104$. This allows us to blindly sample the whole 6--10~keV energy band and search for possible alternative \emph{Xstar} solutions. The $\chi^2$ of the fits plotted against the absorber redshift is reported for all observations in Fig.~12 in Appendix B. This is the best way to derive a physically self-consistent identification of the absorption features. In fact, \emph{Xstar} simultaneously takes into account all the resonance absorption lines and edges from all elements up to $Z$$=$30 for different ionization states. The detailed description of the \emph{Xstar} fits for each observation is discussed in Appendix A and the summary of the best-fit \emph{Xstar} parameters is reported in Table~3.

The possible presence of different minima in the $\chi^2$ distribution indicates that more \emph{Xstar} solutions are possible for distinct redshifts or turbulent velocities. In particular, there are 6/28 cases with two redshift solutions with equivalent $\chi^2$ levels, indicating a possible degeneracy in the identification of the absorption line(s) as resonances from Fe~XXV or Fe~XXVI. In these cases, we considered the best-fit model to be the one associated with the lowest $\chi^2$ value and the relative fit statistic and F-test probability are reported in Table~3. Then, if the possible alternative solutions had $\chi^2$ values consistent at the 90\% level with respect to the best-fit, i.e. with a difference of $\Delta\chi^2$$\le$2.7, we averaged the values of the parameters among them and derived the errors as half of the interval between the lowest and the highest possible values. This allows us to take into account both the statistical errors of measure and the uncertainty in the determination of the best-fit model or, equivalently, of the line identification. However, it is important to note that all the detected narrow Fe K lines are indeed consistent with an identification as resonance transitions essentially from Fe~XXV and Fe~XXVI ions. This is also demonstrated by the overall consistency of the outflow velocities of the absorbers reported in Table~3 and in Table~A.2 of TA. We note that in all the observations only one single \emph{Xstar} component was required to properly model the Fe K absorption lines. 

In Fig.~6, Fig.~7 and Fig.~8 we report the source and background spectra and the best-fit model for three representative cases. More details on the spectral fitting and the best-fit parameters are reported in Appendix A and Table 3 for all the observations, respectively.
Fig.~6 shows the case of PG~1211$+$143 (0112610101) in which a UFO with $v_{out}$$=$$45,300\pm900$~km/s ($0.151\pm0.003$c) has been detected with a high confidence of $P_F$$>$99.99\%. The detection in this observation was already reported by Pounds et al.~(2003). The line is resolved and the velocity width is $\sigma_v$$=$$4,500^{+1,500}_{-1,200}$~km/s. Thus, the absorber best-fit is provided by an \emph{Xstar} grid with turbulent velocity of 5,000~km/s. As it can be seen from Fig.~12 in Appendix B, there is only one solution, which indicates the absorption is mainly due to Fe~XXV and intermediate states Fe~XX--XXIV transitions. The estimated column density and ionization are $N_H$$=$$(8.0^{+2.2}_{-1.1})\times 10^{22}$~cm$^{-2}$ and log$\xi$$=$$2.87^{+0.12}_{-0.10}$~erg~s$^{-1}$~cm, respectively. These values are consistent with the results in Fig.~2. In fact, we can note that the high associated absorption line EW of $\sim$130~eV (see Table~A.2 of TA) for the Fe~XXV He$\alpha$ transition can be reached with an ionization of log$\xi$$\sim$2.9~erg~s$^{-1}$~cm for $N_H$$\sim$$10^{23}$~cm$^{-2}$ and a velocity broadening of $\sim$5,000~km/s. From Fig.~3 we see that for the estimated ionization parameter and velocity broadening, the Fe~XXVI Ly$\alpha$ is $\sim$10 times weaker than the Fe~XXV He$\alpha$ and this explains why it is not observable in the spectrum. Finally, from Fig.~4 we can see that for a velocity width of $\sim$5,000~km/s, the Fe K lines are not saturated up to column densities of $\sim$$10^{23}$--$10^{24}$~cm$^{-2}$. Due to the relatively low ionization level, in the best-fit model in Fig.~6 we can note the presence of additional weak absorption structures at high energies due to the several resonance series converging to the Fe~XX--XXV edges (e.g., Kallman et al.~2004).

Fig.~7 shows the case of Mrk~79 (0400070201) in which we detected a UFO with $v_{out}$$=$$27,600\pm1,200$~km/s ($0.092\pm0.004$c) with a high confidence of $P_F$$=$99.97\%. The relative absorption line is not resolved, but we can place an upper limit of $\sigma_v$$<6,400$~km/s. From Fig.~12 we can note that there is only one redshift solution, consistent with an identification as mainly absorption from Fe~XXVI. As discussed in Appendix A, the best fit is provided by an \emph{Xstar} grid with turbulent velocity of 1,000~km/s (see middle panel in Fig.~7) but the one with 3,000~km/s is equivalent at 90\%. The column density is  $N_H$$=$$(19.4\pm12.0)\times 10^{22}$~cm$^{-2}$ and ionization log$\xi$$=$$4.19\pm0.23$~erg~s$^{-1}$~cm. The EW of the associated blue-shifted absorption line detected by TA is $\sim$40eV. Consistently with Fig.~2, this can be reached with $N_H$$\sim$$10^{23}$~cm$^{-2}$ and log$\xi$$\sim$4~erg~s$^{-1}$~cm for a velocity width of $\sim$1,000--3,000~km/s. From the ratios in Fig.~3 we note that for these values the Fe~XXVI Ly$\alpha$ is indeed more intense, $\sim$5--10 times, than the Fe~XXV He$\alpha$ and this is consistent with the fact that only one absorption line could be detected at high significance given the limited S/N of the spectrum. Moreover, from the ratios in Fig.~4 we see that for velocity widths of $\sim$1,000--3,000~km/s, the 1s--2p lines start to be saturated at $N_H$$\sim$$10^{23}$~cm$^{-2}$, instead the other 1s--3p transitions at higher energies can still grow and their intensities are a bit higher than expected from the simple ratio between the relative oscillator strengths. 

Finally, in Fig.~8 we show the case of NGC~3516 (0401210601) in which we detected a series of lines indicating the presence of a lower velocity system with $v_{out}$$=$$3,300\pm600$~km/s ($0.011\pm0.002$c) with a high confidence of $P_F$$>$99.99\%. The complex absorption in this source was already discussed by Turner et al.~(2008). The lines are unresolved, but the upper limit on the velocity width is $\sigma_v$$<$$2,600$~km/s. From Fig.~12 and Appendix A we can note only one best-fit solution consistent with absorption mainly from Fe~XXV--XXVI.  The best-fit is provided by an \emph{Xstar} grid with turbulent velocity fo 1,000~km/s. The column density is $N_H$$=$$(6.7^{+2.3}_{-1.4})\times 10^{22}$~cm$^{-2}$ and ionization log$\xi$$=$$4.05^{+0.09}_{-0.08}$~erg~s$^{-1}$~cm. The EW of the associated four absorption lines detected by TA is $\sim$25--30~eV and the estimated parameters are quantitatively consistent with what is shown in Fig.~2. Given these parameters, we can see from Fig.~3 and Fig.~4 that the 1s--2p lines are slightly saturated and their ratios with respect to the 1s--3p is slightly lower than what simply expected from their relative oscillator strengths.

\subsection{Consistency checks of the spectral analysis and modeling}

The detailed treatment of the continuum modeling complexities, the background checks, the consistency check with the simultaneous \emph{XMM-Newton} EPIC Metal Oxide Semi-conductor (MOS) camera and the detection significance of the lines through the F-test and extensive Monte Carlo simulations were already reported in TA. As shown in Fig.~6, Fig.~7 and Fig.~8, here we note again that  the background level of the observations of the sample in the 7--10~keV band is always less than $\sim$10\% of the source counts. We also checked that none of the reported blue-shifted absorption lines were induced by an erroneous subtraction of the EPIC pn instrumental background 
emission lines from Ni K$\alpha$ and Cu K$\alpha$ at the energies of 7.48~keV and 
8.05~keV, respectively (Katayama et al.~2004). In fact, only 4/22 of the absorption lines 
detected at E$>$7~keV may have observed energies consistent at the 90\% level with these 
and the features are present in the source spectra with and without background subtraction.
Moreover, the detected features can not be attributed to some sort of EPIC pn calibration 
artifact because they have been detected always at different energies and are narrow. 
Instead, if due to instrumental calibration problems, they would have been expected to be 
present always at some specific energies or, if due to effective area calibration problems, to 
induce broad continuum distortions.

Here, we perform additional tests to check the consistency of the photo-ionization modeling. 
We start with the SED. We checked that a different choice of the power-law slope in the wide range $\Gamma$$=$1.5--2.5 has $<$10\% effects on the $N_H$ and log$\xi$ estimates, within the measurement errors. This is consistent with what derived in \S2.2. 
We then compared the results using the average phenomenological SED derived in \S3.1 with the power-law of $\Gamma$$=$2 assumed in the curve of growth analysis of \S2.2.
We reproduced the EW distributions reported in Fig.~2 for these two cases and find a slight shift toward higher ionization parameters for the average SED to obtain the same value of the power-law SED at the peak.
This effect is essentially due to the different high energy part of these two SEDs. 
In fact, the power-law and average SEDs have the same slope of $\Gamma$$=$2 until E$=$$10^5$~eV, then the average SED drops to zero and the power-law one instead extends up to the upper limit of the \emph{Xstar} input energy band of E$=$$10^6$~eV.
This means that, given the same bolometric luminosity, the average SED has slightly less high energy photons that can effectively ionize iron with respect to the power-law one.
Therefore, to obtain the same iron ionic populations, the average SED needs to increase the global ionization level of the material. 
However, the presence of the high energy cut-off at E$\simeq$100~keV in the average SED causes only a negligible difference in the estimated ionization parameters in Table~3, being less than 5\% and within the measurement errors. Moreover, this high energy cut-off is naturally expected from comptonization models of the X-ray continuum in AGNs (e.g., Haardt \& Maraschi 1991 and references therein) and is in agreement with what is observed for local Seyfert galaxies (e.g., Dadina 2008 and references therein). 
Moreover, as already noted by McKernan et al.~(2003), the presence or absence of the possible UV-bump in the SED has a negligible effect on the parameters of the photo-ionized gas in the Fe K band. 
Again, this and the previous results are essentially due to the fact that only the high energy part of the X-ray continuum at E$\ga$6~keV is capable of photoionizing the highly ionized absorbers studied here, which are dominated by Fe XXV/XXVI.
Therefore, we can directly compare the \emph{Xstar} results in Table~3 with the general trends reported in the curve of growth analysis of \S2.

Another important parameter of the \emph{Xstar} grids is the turbulent velocity of the plasma. It should be noted that the turbulent velocity considered in the grid is different from the velocity broadening parameter $b$ considered in the previous curve of growth analysis (\S2.2), which is the square root of the quadratic sum of the turbulent and thermal velocities. \emph{Xstar} already incorporates the thermal broadening of the lines. However, as already stated in \S2.2, the thermal velocity term for the highly ionized absorbers considered here is essentially negligible and the velocity broadening parameter $b$ is dominated by the turbulent component. 
Therefore, for the highly ionized Fe K features studied here and if not strongly affected by line blending, the velocity parameter $b$ defined in \S2.2, the turbulent velocity of the \emph{Xstar} grids $v_t$ and the velocity width of the lines $\sigma_v$ measured from the spectra are all comparable. We tested that for lower/higher choices of this parameter, in the wide range 
$\sim$100--5000~km/s, the resultant estimate of the ionization parameter is essentially
not affected, although the derived absorber column density is found to be slightly higher/lower. 
This is due to the line saturation effect, which changes the slope of the curve of growth for high column densities, see Fig.~1. 
To compensate for these possible systematics, we used an \emph{Xstar} grid with a turbulent velocity closest to the line velocity width, for the few cases in which the latter was constrained from the data. Instead, for the other cases with only upper limits, we averaged the values of the parameters derived from fits using grids with different turbulent velocities, see \S3.2 and Appendix A.

\section{Results}

The summary of the best-fit parameters of our photo-ionization modeling of the Fe K absorbers detected in the \emph{XMM-Newton} EPIC pn spectra of TA is reported in Table~3. 
We can see that the absorbers are systematically outflowing and their velocities are determined with high accuracy, with errors taking into account also the uncertainty on the line identification (see \S~3.2 and Appendix A). The velocity values are consistent overall with those derived in TA using inverted Gaussians to model the absorption lines and a more simplistic line identification, see \S4.1 and Table~A.2 of TA. The observations in which we detected a UFO are marked with an asterisk in Table~3.

The fraction of sources showing UFOs with respect to the whole sample is again 15/42, which corresponds to $\sim$35\%. However, as already discussed in \S4.2 of TA, if we take into account the fact that we have only conservative estimates on the velocities and that there is a number of spectra in the sample with low S/N, the fraction of UFOs is more probably in the range $\sim$40\%--60\%. The fraction of sources in the sample with UFOs having velocities $\ga$0.1c is $\sim$11/42 ($\sim$25\%). This frequency is instead $\sim$11/15 ($\sim$70\%) only considering those with UFOs. The fraction of sources showing UFOs with respect to the total number with detected Fe K absorbers is 15/19 ($\sim$80\%). Therefore, the great majority of sources with detected Fe K absorbers show the presence of UFOs. The number of observations with detected Fe K absorption lines consistent with UFOs is 19/28 ($\sim$70\%), as reported in Table~3.

The histograms representing the distribution of the outflow velocity, ionization and column density for the UFOs and the lower velocity systems are shown in Fig.~9, Fig.~10 and Fig.~11, respectively. Following the same considerations of TA, we averaged the values reported in Table~3 among the observations of each source and considered the fractions with respect to the total number of sources in the sample (42). Instead, using each observation separately would have introduced a bias because the sources with more observations would have had a higher weight. From Fig.~9 we can see that the velocity distribution of UFOs spans from $\sim$10,000~km/s ($\sim$0.033c) up to $\sim$100,000~km/s ($\sim$0.3c), and has a peak and mean value at $\sim$42,000~km/s ($\sim$ 0.14c). Instead, by definition, the distribution of the lower velocity systems is narrow and has a mean value of $\sim$3,000~km/s ($\sim$0.01c).

However, it should be noted that the present data are affected by several selection effects that hampers the detection of absorption lines at higher energies, or equivalently, of absorbers with higher outflow velocities. An estimate of the magnitude of this effect was already reported in \S4.2 and Fig.~6 of TA. 
This is due to a combination of different causes, such as: the upper limit of the EPIC pn detector bandwidth at $\sim$10~keV, which does not allow the detection of Fe K lines with blueshifts higher than $\sim$0.3--0.4c for the local sources studied here, the drop in effective area and energy resolution of the instrument at E$>$7~keV, the intrinsic lower number of counts for the power-law spectra at high energies and the relatively short exposures. We can derive an estimate of the influence of the limited S/N of the XMM-Newton observations on the derived UFO velocity distribution normalizing the fraction of observations for each velocity interval in the histogram taking into account the number of observations that actually offer enough statistics to detect an absorption line at $\ge 3\sigma$. As already discussed in \S4.2 of TA, this number decreases rapidly for lines with lower EW and higher associated velocity. Consequently, the allowed fraction of observations with higher velocity increases. Assuming the typical absorption line EWs of 20--50~eV, this implies that the peak and mean of the outflow velocity distribution of UFOs in Fig.~9 can potentially shift from $\sim$0.14c up to $\sim$ 0.16--0.18c. 

As reported in Table~3, we have been able to measure the ionization parameter of the absorbers with good accuracy for all observations with Fe K absorbers, taking also into account the uncertainty in the line identification. The distributions of the ionization parameter of the Fe K absorbers ascribable to UFO and to the lower velocity systems are shown in Fig.~10. The values log$\xi$$\sim$2.5 and 6 erg~s$^{-1}$~cm delimit the range in which Fe~XXV/XXVI lines have observable EW$\ga$10~eV, consistently with Fig.~2. We can note that the ionization distribution for the UFOs is much broader than for the other lower velocity absorbers, spanning the whole range from log$\xi$ of 2.5 up to 6 erg~s$^{-1}$~cm, with a mean value of log$\xi$$\sim$4.2~erg~s$^{-1}$~cm. The very high log$\xi$ values of the UFOs may possibly suggest that in some cases the material could not be detected in absorption due to the fact that even iron was completely ionized. Instead, the lower velocity Fe K absorbers are limited in the range log$\xi$$\sim$2.5--4.5~erg~s$^{-1}$~cm, with a mean value of log$\xi$$\sim$3.5~erg~s$^{-1}$~cm.

Fig.~11 shows the distribution of the column densities for the UFOs and the lower velocity systems. We included only the best-fit values in Table~3 for the cases in which the $N_H$ was constrained. We limited the treatment to Compton-thin absorbers with $N_H$$\le$$10^{24}$~cm$^{-2}$. We can see that both distributions have column densities in the wide range $N_H$$\sim$$10^{22}$--$10^{24}$~cm$^{-2}$, with those of the UFOs slightly higher than the lower velocity systems. Their mean values are of $N_H$$\sim$$1.3 \times 10^{23}$~cm$^{-2}$ and $N_H$$\sim$$6.5 \times 10^{22}$~cm$^{-2}$, respectively. This might suggest the presence of possible Compton thick material, at least in the inner regions of the flow.

Finally, it should be noted that the velocities and column densities of the Fe K absorbers derived by fitting the X-ray spectra depend on the unknown inclination angle of the outflows with respect to our line of sight. In other words, they depend whether we are actually looking directly down the outflowing stream or if we intercept only part of it (e.g., Elvis 2000). Therefore, these measures should be regarded only as conservative estimates and the intrinsic values might be higher.

\section{Conclusions}

In this paper we performed an extension of the work presented by us in TA, in which we showed the detection of blueshifted Fe K absorption lines ascribable to UFOs in $\ga$35\% of 42 local radio-quiet AGNs observed with \emph{XMM-Newton}. 
Here, we started with a curve of growth analysis of the main Fe XXV/XXVI K-shell transitions in photo-ionized plasmas. We studied the dependence of the equivalent width of the absorption lines as a function of the ionization parameter, column density, velocity broadening and the lines ratios.
Then, we estimated an average SED for the sample sources and directly modeled the Fe K absorbers in the \emph{XMM-Newton} EPIC pn spectra with the photo-ionization code \emph{Xstar}. A single absorption component is capable of self consistently parameterizing the simple Gaussian absorption lines of TA for each observation. We derived the absorbers column density, ionization parameter and outflow velocity. The summary of the best-fit parameters is reported in Table~3.

We confirm that the frequency of sources showing UFOs in the radio-quiet sample is $\ga$35\%  and that the majority of the Fe K absorbers are indeed associated with UFOs ($\ga$80\%). We derived that the outflow velocity distribution of the UFOs spans from $\sim$10,000~km/s ($\sim$0.03c) up to $\sim$100,000~km/s ($\sim$0.3c), with a peak and mean value at $\sim$42,000--55,000 km/s ($\sim$0.14--0.18c), taking into account also the presence of selection effects that hampers the detection of the higher velocity absorbers. These effects can only be reduced by performing much longer observations of these sources with the present instruments or using the higher sensitivity and energy resolution of the calorimeters on board the future Astro-H and the proposed Athena missions.

The ionization parameter of the UFOs is very high and in the wide range log$\xi$$\sim$3--6~erg~s$^{-1}$~cm, consistent with Fe~XXV/XXVI being the dominant lines, with a mean value of log$\xi$$\sim$4.2~erg~s$^{-1}$~cm. This might give rise to the possibility of the presence of additional columns of material, which are unseen to absorption line spectroscopy because even iron is completely ionized.
The column densities associated with the UFOs are also large and in the broad range $N_H$$\sim$$10^{22}$--$10^{24}$~cm$^{-2}$, with a mean value of $N_H$$\sim$$10^{23}$~cm$^{-2}$. This might suggest the presence of Compton thick material, at least in the inner regions of the flow.
However, besides possible selection effects, the measured velocities and column densities should also be regarded as conservative estimates because they depend on the unknown inclination angle of the flow with respect to the line of sight.

Depending on the actual outflow geometry and covering factor, the latter estimated from the detection frequency on the whole sample of $\sim$0.4--0.6 in \S5 of TA, this material might be observed also in emission, through fluorescence/recombination lines, continuum reflection, bremsstrahlung and/or black body emission. The electron temperature of kT$\sim$100~eV (see \S2.2) then suggests the continuum emission might be mainly in the soft X-rays (e.g., King \& Pounds 2003; Pounds \& Reeves 2007; Sim et al.~2008; Sim et al.~2010 and references therein).
Finally, it is important to note that the parameters here derived for the UFOs in the local radio-quiet AGN sample of TA in Table~3 are consistent overall with those estimated for the complementary, but much smaller, local radio-loud AGN sample in TB, suggesting a possible link between the two AGN classes.

We refer to a successive paper III of this series (Tombesi et al.~in prep.) for a detailed discussion of the physical properties, geometry and energetics of the UFOs here analyzed and also a comparison with the classical AGN warm absorbers. Finally, significant improvements are expected from the high effective area and supreme energy resolution in the Fe K band offered by the X-ray microcalorimeters on board the future Astro-H and the proposed Athena missions.

\acknowledgments

FT thank T.~R. Kallman for the useful discussions on the use of the \emph{Xstar} code. FT thanks T. Yaqoob and the Johns Hopkins University for the visiting period spent there doing part of this work. FT thank C.~S. Reynolds for useful discussions. This paper is based on observations obtained with the \emph{XMM-Newton} satellite, an ESA funded mission with contributions by ESA member states and USA. FT acknowledge support from NASA through the ADAP/LTSA program. This research has made use of the NASA/IPAC Extragalactic Database (NED) which is operated by the Jet Propulsion Laboratory, California Institute of Technology, under contract with the National Aeronautics and Space Administration. MC, GP and MD acknowledge support from ASI under the contract INAF/ASI I/088/06/0. VB acknowledge support from the UK STFC research council. The authors thank the anonymous referee for suggestions that led to important improvements in the paper.

\clearpage

   \begin{figure}[!t]
   \centering
    \includegraphics[width=8.5cm,height=14cm,angle=0]{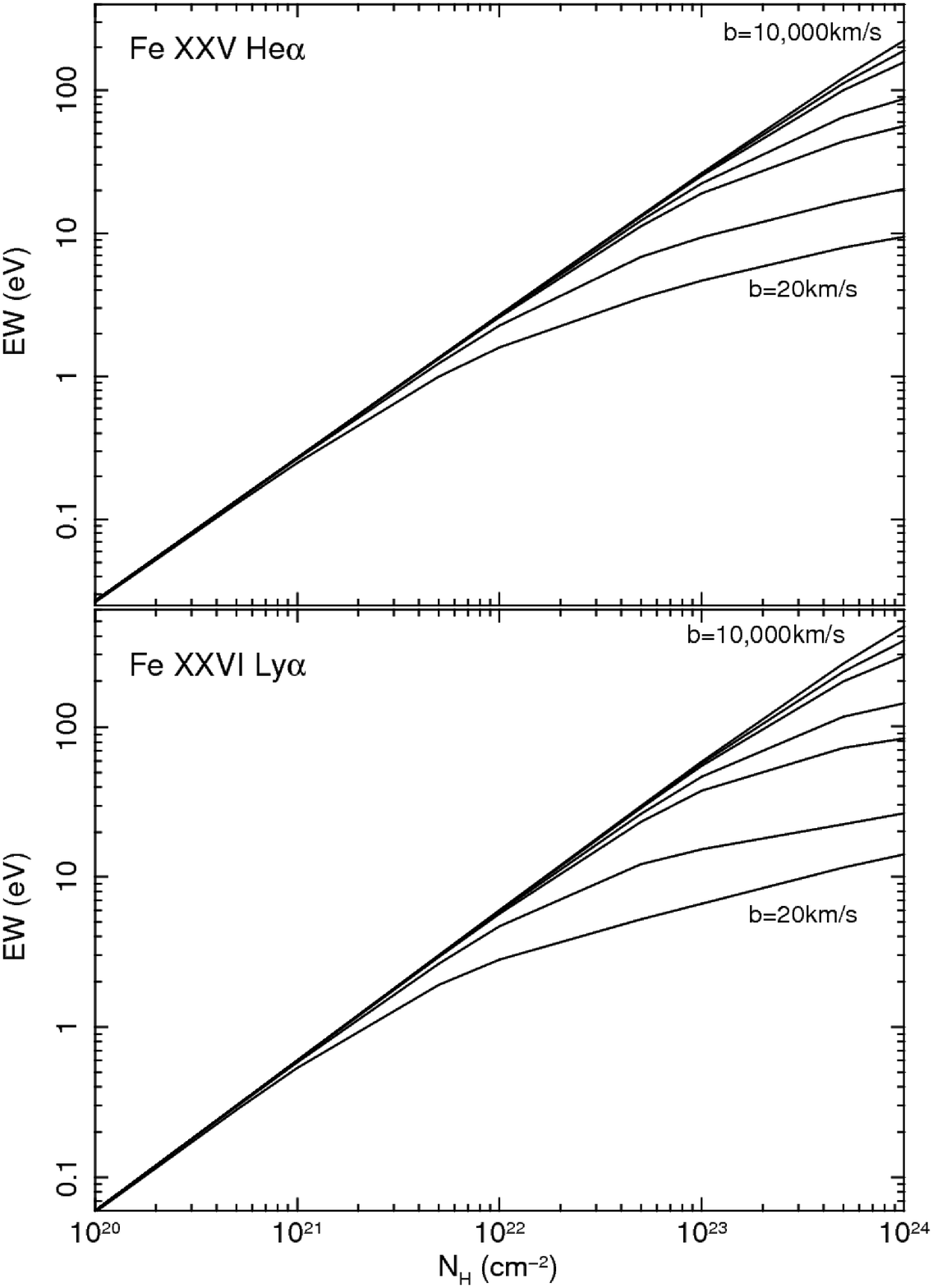}
   \caption{Curve of growth for the Fe XXV He$\alpha$ (r $+$ i) (\emph{upper panel}) and Fe XXVI Ly$\alpha$ (r$_1$ $+$ r$_2$) (\emph{lower panel}) lines, for different values of the velocity broadening parameter $b$. From bottom to top $b$$=$20, 100, 500, 1000, 3000, 5000, 10000~km/s. The ionization parameter is fixed to log$\xi$$=$3.5~erg~s$^{-1}$~cm.}
    \end{figure}

   \begin{figure}[!t]
   \centering
    \includegraphics[width=8.5cm,height=14cm,angle=0]{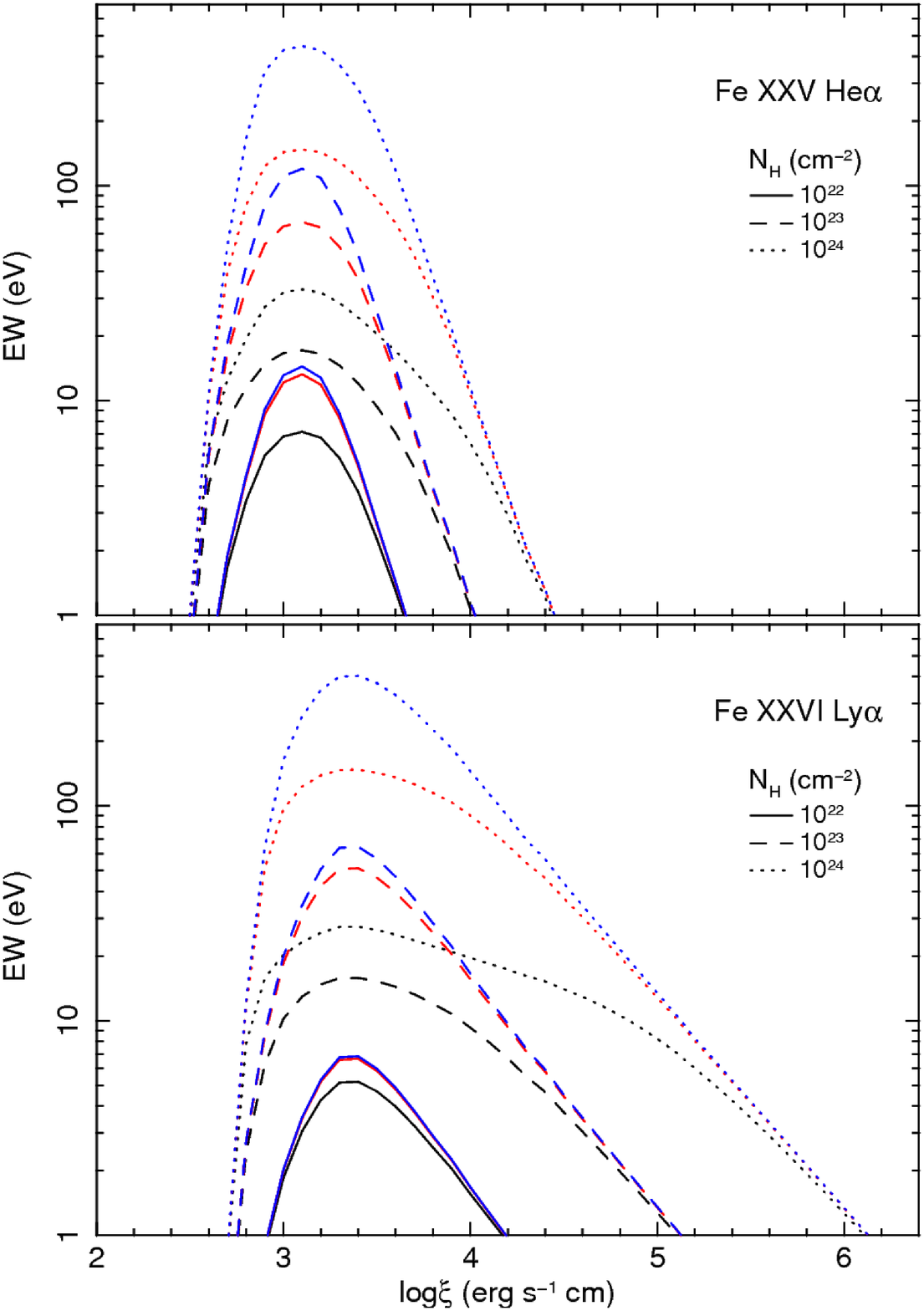}
   \caption{Estimated EW of the Fe XXV He$\alpha$ (r $+$ i) (\emph{upper panel}) and Fe XXVI Ly$\alpha$ (r$_1$ $+$ r$_2$) (\emph{lower panel}) transition as a function of log$\xi$, for different column densities. The velocity broadening is $b$$=$100 (black), 1000 (red), 5000~km/s (blue).}
    \end{figure}

   \begin{figure}[!t]
   \centering
    \includegraphics[width=7cm,height=8.5cm,angle=270]{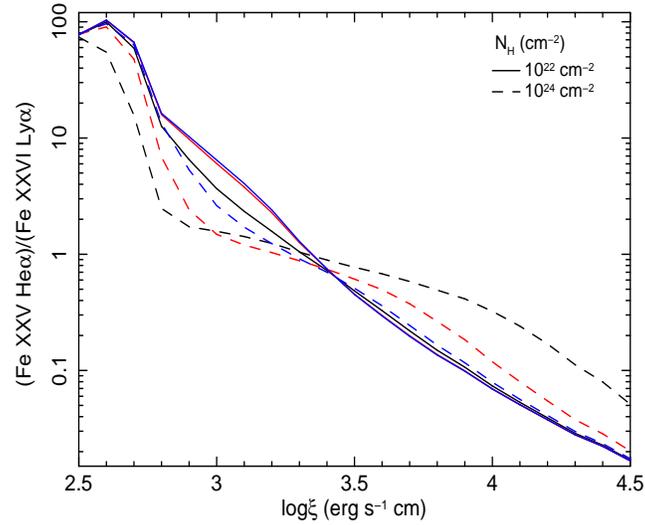}
   \caption{Ratio of the EWs of the Fe XXV He$\alpha$ (r $+$ i) and Fe XXVI Ly$\alpha$ (r$_1$ $+$ r$_2$) lines as a function of log$\xi$, for different column densities. The velocity broadening is $b$$=$ 100 (black), 1000 (red), 5000~km/s (blue).}
    \end{figure}

   \begin{figure}[!t]
   \centering
    \includegraphics[width=8.5cm,height=14cm,angle=0]{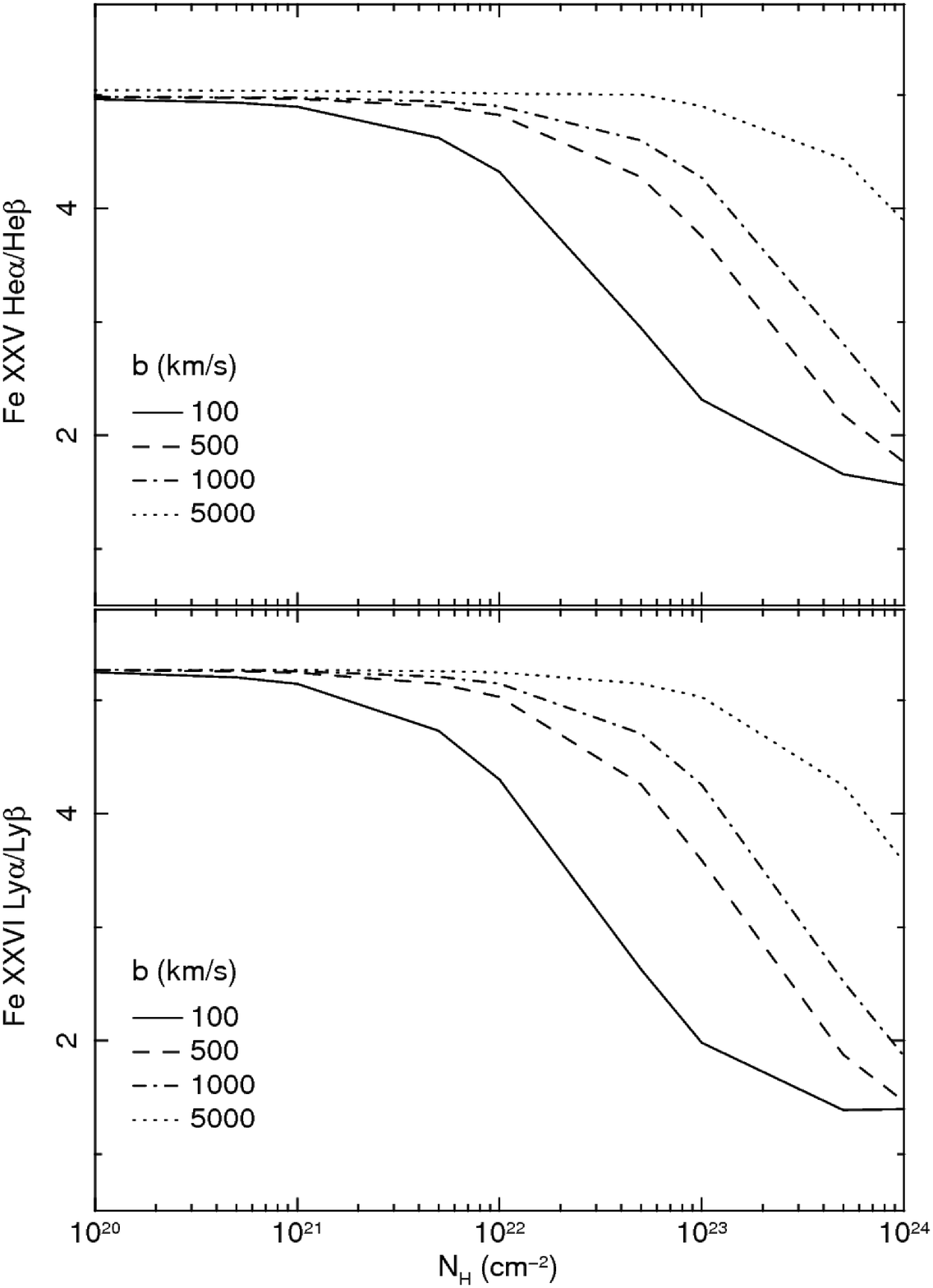}
   \caption{Ratio of the EWs of the Fe XXV He$\alpha$/He$\beta$ (r $+$ i) (\emph{upper panel}) and Fe XXVI Ly$\alpha$/Ly$\beta$ (r$_1$ $+$ r$_2$) (\emph{lower panel}) lines as a function of the column density, for different values of the velocity parameter $b$. The ionization parameter is fixed to log$\xi$$=$3.5~erg~s$^{-1}$~cm.}
    \end{figure}

   \begin{figure}[!t]
   \centering
    \includegraphics[width=7cm,height=8.5cm,angle=270]{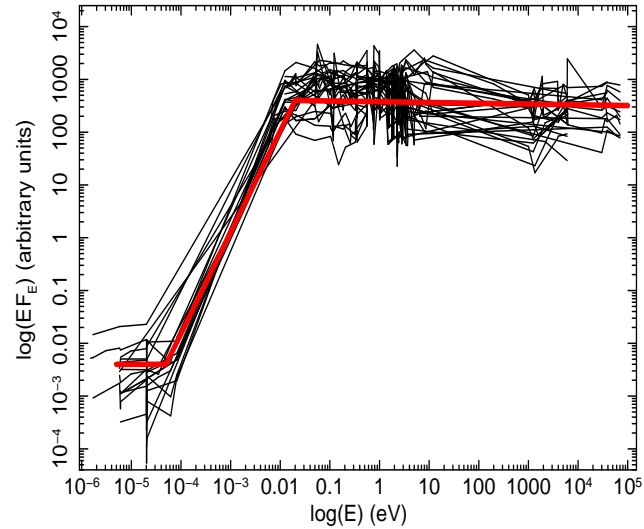}
   \caption{SEDs of the radio-quiet type 1 AGNs in the sample normalized at the near-IR inflection point (black lines) and the resultant average SED (red line).}
    \end{figure}

   \begin{figure}[!t]
   \centering
    \includegraphics[width=8.8cm,height=11.3cm,angle=0]{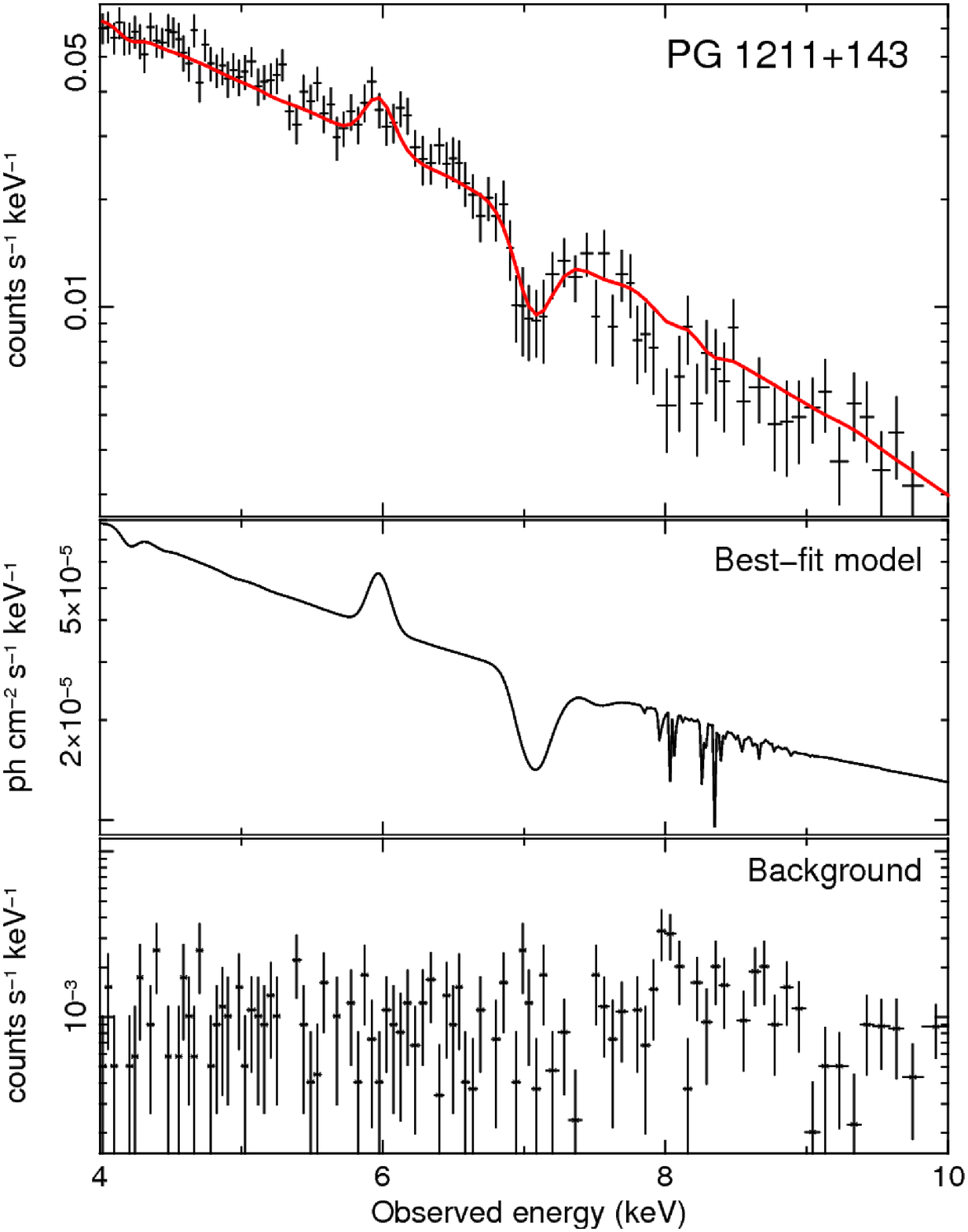}
   \caption{\emph{XMM-Newton} EPIC pn spectrum of PG~1211$+$143 (0112610101) in the 4--10~keV band in which a UFO with $v_{out}$$\sim$0.15c has been detected. \emph{Upper panel:} background subtracted source spectrum with superimposed the best-fit model. \emph{Middle panel:} best-fit model using an \emph{Xstar} grid with turbulent velocity of 5000~km/s. \emph{Lower panel:} background spectrum. Data are binned to 1/5 of the FWHM of the instrument.}
    \end{figure}

   \begin{figure}[!t]
   \centering
    \includegraphics[width=8.8cm,height=11.3cm,angle=0]{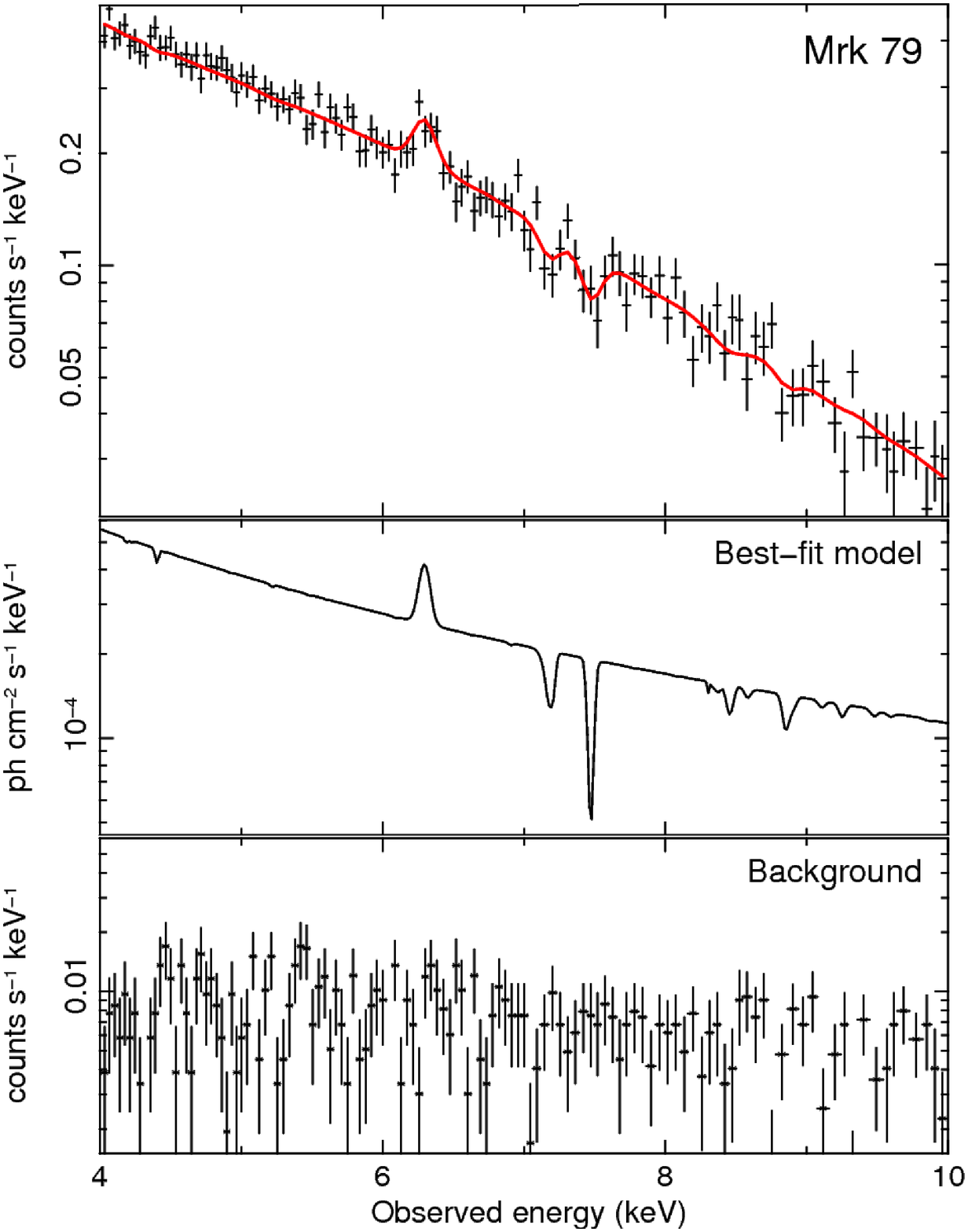}
    \caption{\emph{XMM-Newton} EPIC pn spectrum of Mrk~79 (0400070201) in the 4--10~keV band in which a UFO with $v_{out}$$\sim$0.1c has been detected. \emph{Upper panel:} background subtracted source spectrum with superimposed the best-fit model. \emph{Middle panel:} best-fit model using an \emph{Xstar} grid with turbulent velocity of 1000~km/s. \emph{Lower panel:} background spectrum. Data are binned to 1/5 of the FWHM of the instrument.}
    \end{figure}

   \begin{figure}[!t]
   \centering
    \includegraphics[width=8.8cm,height=11.3cm,angle=0]{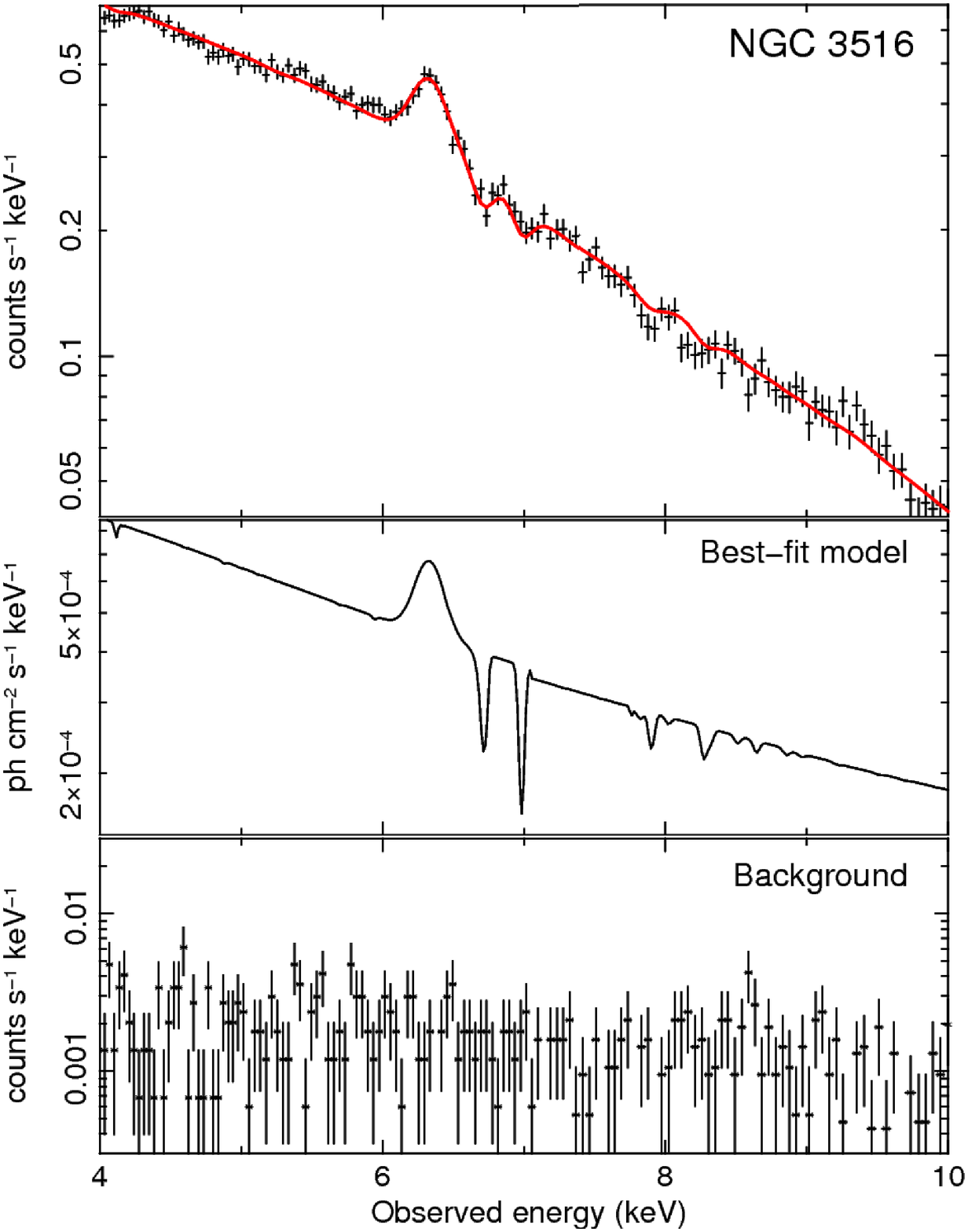}
   \caption{\emph{XMM-Newton} EPIC pn spectrum of NGC~3516 (0401210601) in the 4--10~keV band in which a lower velocity system with $v_{out}$$\sim$0.01c has been detected. \emph{Upper panel:} background subtracted source spectrum with superimposed the best-fit model. \emph{Middle panel:} best-fit model using an \emph{Xstar} grid with turbulent velocity of 1000~km/s. \emph{Lower panel:} background spectrum. Data are binned to 1/5 of the FWHM of the instrument.}
    \end{figure}

   \begin{figure}[!t]
   \centering
    \includegraphics[width=7cm,height=8.5cm,angle=270]{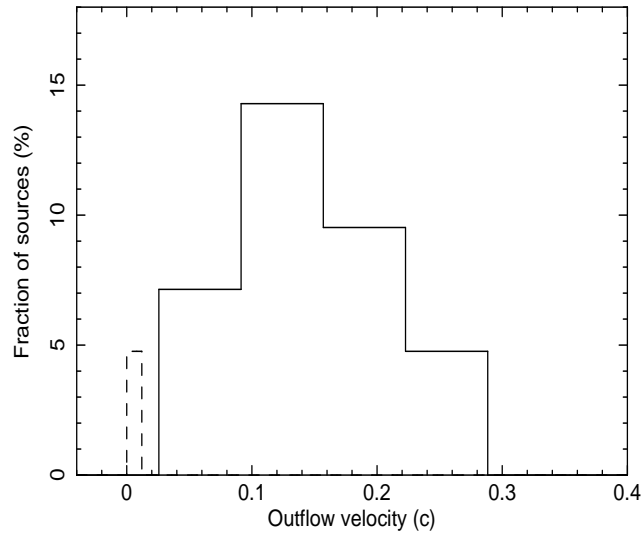}
    \caption{Histograms representing the distribution of the outflow velocity of UFOs (solid line) and the lower velocity systems (dashed line), normalized to the total number of sources in the sample.}
    \end{figure}

   \begin{figure}[!t]
   \centering
    \includegraphics[width=7cm,height=8.5cm,angle=270]{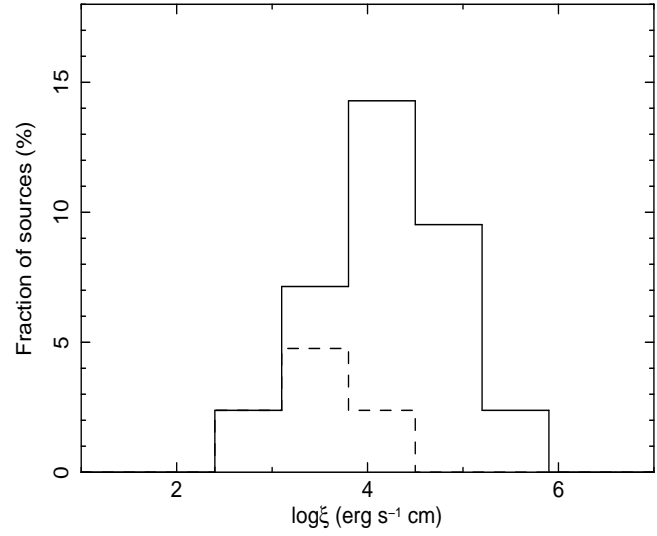}
   \caption{Histograms representing the distribution of the ionization parameter of the as UFOs (solid line) and the lower velocity systems (dashed line), normalized to the total number of sources in the sample.}
    \end{figure}

   \begin{figure}[!t]
   \centering
    \includegraphics[width=7cm,height=8.5cm,angle=270]{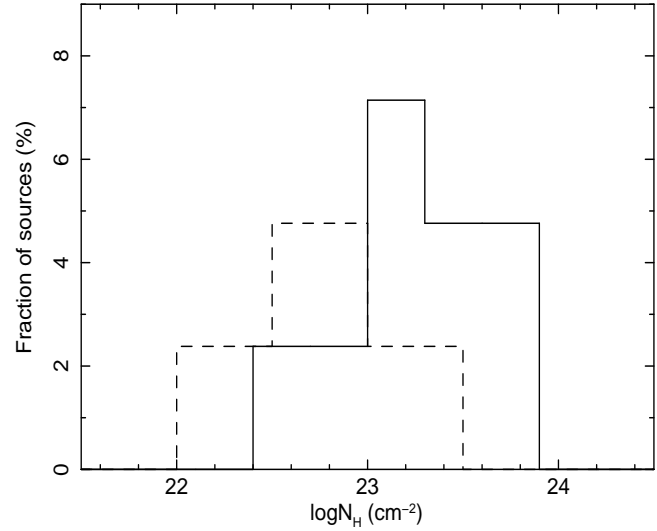}
   \caption{Histograms representing the distribution of the column density of the UFOs (solid line) and lower velocity systems (dashed line), normalized to the total number of sources in the sample.}
    \end{figure}

\clearpage

\begin{deluxetable}{l c c c c c r r}
\tabletypesize{\scriptsize}
\tablecaption{Parameters of the considered Fe XXV and Fe XXVI K-shell transitions.}
\tablewidth{0pt}
\tablehead{\colhead{Ion} & \colhead{ID} & \colhead{Transition} & \colhead{$<$E$>$} & \colhead{Line} & \colhead{E} & \colhead{$f_{lu}$} & \colhead{$A_{ul}$}\\ 
& & & \colhead{\scriptsize{keV}} & & \colhead{\scriptsize{(keV)}} & \colhead{\scriptsize{($\times 10^{-2}$)}} & \colhead{\scriptsize{($\times 10^{12}$)}}}
\startdata
Fe XXV & He$\alpha$ & 1s$^2$--1s2p & 6.697 & (r) & 6.700 & 70.4 & 457\\
 & & & & (i) & 6.668 & 6.9 & 44\\
       & He$\beta$ &  1s$^2$--1s3p& 7.880 & (r) & 7.881 & 13.8 & 124\\
 & & & & (i) & 7.872 & 1.7 & 1\\
Fe XXVI & Ly$\alpha$ & 1s--2p & 6.966 & (r$_1$) & 6.973 & 28.0 & 296\\
 & & & & (r$_2$) & 6.952 & 14.0 & 293\\
        & Ly$\beta$ & 1s--3p & 8.250 & (r$_1$) & 8.253 & 5.3 & 79\\
 & & & & (r$_2$) & 8.246 & 2.6 & 78\\
\enddata
\tablecomments{The line parameters have been taken from the accurate NIST atomic database. (1) Iron ion. (2) Line identification. (3) Atomic transition. (4) Average line energy, weighted for the oscillator strengths of its components. (5) Line components. (6) Energy of the specific transition. (7) Oscillator strength. (8) Einstein coefficient.}
\end{deluxetable}

\clearpage

\begin{deluxetable}{c c c c}
\tabletypesize{\scriptsize}
\tablecaption{Average phenomenological SED for the type 1 objects of the TA radio-quiet AGN sample.}
\tablewidth{0pt}
\tablehead{\colhead{Interval} & \colhead{E (eV)} & \colhead{$\Gamma$} & \colhead{$\alpha$}}
\startdata
 1 & $5\times10^{-6}$--$5\times10^{-5}$ & 2 & $-1$ \\      
 2 & $5\times10^{-5}$--$2\times10^{-2}$ & 0.1  & 0.9 \\
 3 & $2\times10^{-2}$--$1\times10^{+5}$ & 2  & $-1$\\
\enddata
\end{deluxetable}

\clearpage

\begin{deluxetable}{lc|c|ccc|ccc}
\tabletypesize{\scriptsize}
\tablecaption{Summary of the best-fit \emph{Xstar} parameters.}
\tablecolumns{9}
\tablewidth{0pt}
\tablehead{\multicolumn{1}{l}{Source} & \colhead{OBSID}  & \colhead{$\sigma_v$} & \colhead{$v_{out}$}  & \colhead{log$\xi$} & \colhead{$N_H$} & \colhead{$\Delta\chi^2/\Delta\nu$} & \colhead{$\chi^2/\nu$} & \colhead{$P_F$}}
\startdata
NGC~4151 & 0402660201\tablenotemark{*} & $5.1^{+1.8}_{-1.4}$  & $0.106^{+0.007}_{-0.007}$ & $4.41^{+0.92}_{-0.08}$  & $>2$ & 21.4/3 & $219.6/157$ & 99.80\%\\[1.5pt]
IC4329A & 0147440101\tablenotemark{*}  & $<3.1$ & $0.098^{+0.004}_{-0.004}$ & $5.34^{+0.94}_{-0.94}$ & $>2$ & 12.5/3 & 240.6/159  & 97.00\% \\[1.5pt]
NGC~3783 & 0112210101  & $<2.6$ & $0.010^{+0.005}_{-0.005}$ & $2.13^{+0.15}_{-0.09}$ & $2.3^{+0.9}_{-0.5}$ & 17.4/3 & 201.6/160 & 99.60\% \\[1.5pt] 
 & 0112210201  & $<2.7$ & $<0.007$ & $2.87^{+0.06}_{-0.05}$ & $2.8^{+0.2}_{-0.2}$ & 40.4/3 & 173.5/157 & $>$99.99\% \\[1.5pt]
 & 0112210501 & $<2.6$ & $<0.007$ & $2.98^{+0.06}_{-0.06}$ & $2.7^{+0.2}_{-0.2}$ & 77.8/3 & 193.8/157 & $>$99.99\% \\[1.5pt]
NGC~3516 & 0401210401  & $<2.7$ & $0.006^{+0.002}_{-0.002}$ & $4.04^{+0.11}_{-0.14}$ & $5.4^{+2.0}_{-1.4}$ & 58.4/3 & 205.4/197 & $>$99.99\% \\[1.5pt]
 & 0401210501  & $<2.8$ & $0.008^{+0.002}_{-0.002}$ & $4.18^{+0.09}_{-0.06}$ & $7.0^{+3.2}_{-1.2}$ & 45.3/3 & 195.0/156 & $>$99.99\% \\[1.5pt]
 & 0401210601  & $<2.6$ & $0.011^{+0.002}_{-0.002}$ & $4.05^{+0.09}_{-0.08}$ & $6.7^{+2.3}_{-1.4}$ & 54.8/3 & 170.3/154 & $>$99.99\% \\[1.5pt]
 & 0401211001  & $<2.8$ & $0.013^{+0.003}_{-0.003}$ & $4.18^{+0.12}_{-0.10}$ & $7.1^{+2.4}_{-2.2}$ & 25.1/3 & 179.6/154 & 99.98\% \\[1.5pt]
Mrk~509 & 0130720101\tablenotemark{*}  & $<3.8$ & $0.173^{+0.004}_{-0.004}$ & $5.06^{+0.58}_{-0.58}$ & $>6$ & 11.8/3 & 139.2/150 & 99.40\%\\[1.5pt]
 & 0306090201\tablenotemark{*}  & $<3.2$ & $0.138^{+0.004}_{-0.004}$ & $5.02^{+0.98}_{-0.98}$ & $>6$ & 12.7/3 & 177.0/158 & 99.00\%\\[1.5pt]
 & 0306090401\tablenotemark{*}  & $<5.0$ & $0.217^{+0.024}_{-0.024}$ & $4.25^{+1.46}_{-1.46}$ & $>0.7$ & 14.5/3 & 201.0/159 & 99.00\%\\[1.5pt]
Ark~120 & 0147190101\tablenotemark{*}  & $<3.3$ & $0.287^{+0.022}_{-0.022}$ & $4.55^{+1.29}_{-1.29}$ & $>0.7$ & 12.5/3 & 149.3/153 & 99.40\% \\[1.5pt]
Mrk~279 & 0302480501 & $<4.5$ & $<0.007$ & $3.73^{+0.59}_{-0.59}$ & $4.6^{+3.6}_{-3.6}$ & 14.0/3 & 132.6/152 & 99.85\% \\[1.5pt]
Mrk~79 & 0400070201\tablenotemark{*} & $<6.4$ & $0.092^{+0.004}_{-0.004}$ & $4.19^{+0.23}_{-0.23}$ & $19.4^{+12.0}_{-12.0}$ & 19.1/3 & 131.9/138 & 99.97\%\\[1.5pt]
NGC~4051 & 0109141401\tablenotemark{*} & $<4.2$ & $0.037^{+0.025}_{-0.025}$ & $4.37^{+1.54}_{-1.54}$ & $>0.8$ & 13.8/3 & 160.5/155 & 99.50\% \\[1.5pt]
 & 0157560101\tablenotemark{*} & $4.7^{+1.8}_{-1.5}$ & $0.202^{+0.006}_{-0.006}$ & $3.06^{+0.17}_{-0.19}$ & $5.2^{+1.1}_{-1.3}$ & 22.0/3 & 152.6/131 & 99.95\%\\[1.5pt] 
Mrk~766 & 0304030301\tablenotemark{*} & $4.6^{+2.7}_{-1.8}$ & $0.082^{+0.006}_{-0.006}$ & $3.46^{+0.71}_{-0.32}$ & $2.1^{+0.4}_{-0.3}$ & 22.4/3 & 182.6/154 & 99.95\%\\[1.5pt]
 & 0304030501\tablenotemark{*}  & $<4.4$ & $0.088^{+0.002}_{-0.002}$ & $4.28^{+0.08}_{-0.11}$ & $10.0^{+6.4}_{-3.5}$ & 21.2/3 & 187.1/151 & 99.90\%\\[1.5pt]
Mrk~841 & 0205340401\tablenotemark{*} & $<4.0$ & $0.055^{+0.025}_{-0.025}$ & $3.91^{+1.24}_{-1.24}$ & $>1$ & 12.7/3 & 134.9/127 & 99.07\% \\[1.5pt]
ESO~323-G77 & 0300240501 & $<3.7$ & $0.012^{+0.004}_{-0.004}$ & $3.74^{+0.37}_{-0.37}$ & $23.4^{+12.3}_{-12.3}$ & 77.0/3 & 146.8/132 & $>$99.99\% \\[1.5pt]
1H0419-577 & 0148000201\tablenotemark{*} & $<7.0$ & $0.079^{+0.007}_{-0.007}$ & $3.69^{+0.53}_{-0.53}$ & $17.3^{+14.3}_{-14.3}$ & 14.8/3 & 113.9/102 & 99.43\% \\[1.5pt]
Mrk~290 & 0400360601\tablenotemark{*}  & $<4.0$ & $0.163^{+0.024}_{-0.024}$ & $3.91^{+1.17}_{-1.17}$ & $25.9^{+24.0}_{-24.0}$ & 13.5/3 & 90.2/101 & 99.74\% \\[1.5pt]
Mrk~205 & 0124110101\tablenotemark{*} & $<3.1$ & $0.100^{+0.004}_{-0.004}$ & $4.92^{+0.56}_{-0.56}$ & $>14.5$ & 11.9/3 & 123.8/118 & 98.80\% \\[1.5pt]
PG~1211+143 & 0112610101\tablenotemark{*} & $4.5^{+1.5}_{-1.2}$ & $0.151^{+0.003}_{-0.003}$ & $2.87^{+0.12}_{-0.10}$ & $8.0^{+2.2}_{-1.1}$ & 70.5/3 & 125.5/114 & $>$99.99\%\\[1.5pt]

MCG-5-23-16 & 0302850201\tablenotemark{*} & $<13.0$ & $0.116^{+0.004}_{-0.004}$ & $4.33^{+0.08}_{-0.08}$ & $4.0^{+1.2}_{-1.2}$ & 41.3/3 & 257.1/158 & $>$99.99\% \\[1.5pt]
NGC~4507 & 0006220201\tablenotemark{*} & $<4.2$ & $0.199^{+0.024}_{-0.024}$ & $4.53^{+1.15}_{-1.15}$ & $>0.9$ & 10.1/3 & 148.7/144 & 97.70\% \\[1.5pt]
NGC~7582 & 0112310201\tablenotemark{*} & $<11.0$ & $0.285^{+0.003}_{-0.003}$ & $3.39^{+0.09}_{-0.15}$ & $23.4^{+21.1}_{-9.8}$ & 20.9/3 & 111.0/100 & 99.94\%\\[1.5pt]
\enddata
\tablecomments{The errors are at the 1$\sigma$ level. The lower limits on $N_H$ and upper limits on $v_{out}$ are at the 90\% confidence level. The $^*$ marks the observations with detected UFO. (1) Source name. (2) \emph{XMM-Newton} observation ID. (3) Absorption lines velocity width in units of $10^3$~km/s. (4) Outflow velocity in units of c. (5) Logarithm of the ionization parameter in units of erg~s$^{-1}$~cm. (6) Absorber column density in units of $10^{22}$~cm$^{-2}$. (7) $\chi^2$ improvement after including the \emph{Xstar} table in the baseline model and additional model parameters. (8) Final best-fit $\chi^2$ after including the \emph{Xstar} table and degrees of freedom. (9) F-test probability.} 
\end{deluxetable}

\clearpage

\appendix

\section{Notes on single sources}

\subsection{NGC~4151}

Observation 0402660201: the velocity width of the absorption line is measured as $\sigma_v$$=$ $5100^{+1800}_{-1400}$~km/s. Fig.~12 in Appendix B shows the fit solution for the \emph{Xstar} grid with $v_t$$=$ 5000~km/s. The parameters are log$\xi$$=$$4.41^{+0.92}_{-0.08}$ erg~s$^{-1}$~cm, $N_H$$>$$2 \times 10^{22}$ cm$^{-2}$ and redshift $z$$=$$-0.099\pm0.007$, which corresponds to $v_{out}$$=$$0.106\pm0.007c$. The initial baseline model statistic is $\chi^2$$=$241.0 for $d.o.f.$$=$160 and the fit improvement is $\Delta \chi^2$$=$21.4 for three model parameters. This provides the best-fit solution, with F-test probability is $P_F$$=$99.8\%. We detected for the first time a UFO in this observation.

\subsection{IC4329A}

Observation 0147440101: the velocity width of the absorption line is $\sigma_v$$<$ 3100~km/s. From Fig.~12 in Appendix B there is one redshift solution. The initial baseline model statistic is $\chi^2$$=$253.1 for $d.o.f.$$=$162. Using an \emph{Xstar} grid with $v_t$$=$1000~km/s we obtain (1) log$\xi$$=$$5.17^{+0.75}_{-0.77}$~erg~s$^{-1}$~cm, $N_H$$>$$2 \times 10^{22}$~cm$^{-2}$ and $z$$=$$-0.078\pm0.003$, which corresponds to $v_{out}$$=$$0.097\pm0.003c$. The fit improvement is $\Delta \chi^2$$=$12.5 for three model parameters. Instead, using an \emph{Xstar} grid with $v_t$$=$3000~km/s we obtain (2) log$\xi$$=$$5.63^{+0.65}_{-1.15}$~erg~s$^{-1}$~cm, $N_H$$>$$2 \times 10^{22}$~cm$^{-2}$, $z$$=$$-0.079\pm0.003$, which corresponds to $v_{out}$$=$$0.098\pm0.003c$. The fit improvement is $\Delta \chi^2$$=$10.8 for three model parameters. The best-fit is provided by (1), with F-test probability $P_F$$=$97\%. The fits (1) and (2) are equivalent at 90\%. The detection of a possible UFO in this observation was already reported by Markowitz et al.~(2006).

\subsection{NGC~3783}

Observation 0112210101: the velocity width of the absorption line is $\sigma_v$$<$ 2600~km/s. From Fig.~12 in Appendix B, there is only one redshift solution. The initial baseline model statistic is $\chi^2$$=$219.0 for $d.o.f.$$=$163. Using an \emph{Xstar} grid with $v_t$$=$1000~km/s we obtain log$\xi$$=$$2.13^{+0.15}_{-0.09}$ erg~s$^{-1}$~cm, $N_H$$=$$(2.3^{+0.9}_{-0.5}) \times 10^{22}$~cm$^{-2}$ and $z$$=$$-0.001\pm0.005$, which corresponds to $v_{out}$$=$$0.010\pm0.005c$. The fit improvement is $\Delta \chi^2$$=$17.4 for three model parameters. This provides the best-fit, with F-test probability $P_F$$=$99.6\%.

Observation 0112210201: the velocity width of the absorption lines is $\sigma_v$$<$ 2700~km/s. From Fig.~12 in Appendix B, there is only one redshift solution. The initial baseline model statistic is $\chi^2$$=$213.9 for $d.o.f.$$=$160. Using an \emph{Xstar} grid with $v_t$$=$1000~km/s we obtain log$\xi$$=$$2.87^{+0.06}_{-0.05}$~erg~s$^{-1}$~cm, $N_H$$=$$(2.8^{+0.2}_{-0.2}) \times 10^{22}$~cm$^{-2}$ and redshift $z$$=$$0.009\pm0.004$, which corresponds to $v_{out}$$<$$0.007c$. $\Delta \chi^2$$=$40.4 for three model parameters. This provides the best-fit, with F-test probability $P_F$$>$99.99\%.

Observation 0112210501: the velocity width of the absorption line is $\sigma_v$$<$ 2600~km/s. From Fig.~12 in Appendix B, there is only one redshift solution. The initial baseline model statistic is $\chi^2$$=$271.6 for $d.o.f.$$=$160. Using an \emph{Xstar} grid with $v_t$$=$1000~km/s we obtain log$\xi$$=$$2.98^{+0.06}_{-0.06}$~erg~s$^{-1}$~cm, $N_H$$=$$(2.7^{+0.2}_{-0.2}) \times 10^{22}$~cm$^{-2}$ and $z$$=$$0.008\pm0.004$, which corresponds to $v_{out}$$<$$0.007c$. The fit improvement is $\Delta \chi^2$$=$77.8 for three model parameters. This provides the best-fit, with F-test probability $P_F$$>$99.99\%. A detection of an Fe K absorber was already reported by Reeves et al.~(2004). Our results reported in Table 3 are consistent overall with theirs.

\subsection{NGC~3516}

Observation 0401210401: the velocity width of the absorption line is $\sigma_v$$<$ 2700~km/s. From Fig.~12 in Appendix B, there is only one redshift solution. The initial baseline model statistic is $\chi^2$$=$263.8 for $d.o.f.$$=$160. Using an \emph{Xstar} grid with $v_t$$=$1000~km/s we obtain log$\xi$$=$$4.04^{+0.11}_{-0.14}$~erg~s$^{-1}$~cm, $N_H$$=$$(5.4^{+2.0}_{-1.4}) \times 10^{22}$~cm$^{-2}$ and $z$$=$$0.003\pm0.002$, which corresponds to $v_{out}$$=$$0.006\pm0.002c$. The fit improvement is $\Delta \chi^2$$=$58.4 for three model parameters. This provides the best-fit, with F-test probability $P_F$$>$99.99\%.

Observation 0401210501: the velocity width of absorption lines is $\sigma_v$$<$ 2800 km/s. From Fig.~12 in Appendix B, there is only one redshift solution. The initial baseline model statistic is $\chi^2$$=$240.3 for $d.o.f.$$=$159. Using an \emph{Xstar} grid with $v_t$$=$1000~km/s we obtain log$\xi$$=$$4.18^{+0.09}_{-0.06}$~erg~s$^{-1}$~cm, $N_H$$=$$(7.0^{+3.2}_{-1.2}) \times 10^{22}$~cm$^{-2}$ and $z$$=$$0.001\pm0.002$, which corresponds to $v_{out}$$=$$0.008\pm0.002c$. The fit improvement is $\Delta \chi^2$$=$45.3 for three model parameters. This provides the best-fit, with F-test probability $P_F$$>$99.99\%.

Observation 0401210601: the velocity width of the absorption lines is $\sigma_v$$<$ 2600~km/s. From Fig.~12 in Appendix B, there is only one redshift solution. The initial baseline model statistic is $\chi^2$$=$225.1 for $d.o.f.$$=$157. Using an \emph{Xstar} grid with $v_t$$=$1000~km/s we obtain log$\xi$$=$$4.05^{+0.09}_{-0.08}$~erg~s$^{-1}$~cm, $N_H$$=$$(6.7^{+2.3}_{-1.4}) \times 10^{22}$~cm$^{-2}$ and $z$$=$$-0.002\pm0.002$, which corresponds to $v_{out}$$=$$0.011\pm0.002c$. The fit improvement is $\Delta \chi^2$$=$54.8 for three model parameters. This provides the best-fit, with F-test probability $P_F$$>$99.99\%.

Observation 0401211001: the velocity width of the absorption lines is $\sigma_v$$<$ 2800~km/s. From Fig.~12 in Appendix B, there is only one redshift solution. The initial baseline model statistic is $\chi^2$$=$204.7 for $d.o.f.$$=$157. Using an \emph{Xstar} grid with $v_t$$=$1000~km/s we obtain log$\xi$$=$$4.18^{+0.12}_{-0.10}$~erg~s$^{-1}$~cm, $N_H$$=$$(7.1^{+2.4}_{-2.2}) \times 10^{22}$~cm$^{-2}$ and $z$$=$$-0.004\pm0.003$, which corresponds to $v_{out}$$=$$0.013\pm0.003c$. The fit improvement is $\Delta \chi^2$$=$25.1 for three model parameters. This provides the best-fit, with F-test probability $P_F$$=$99.98\%. The detection of an Fe K absorber was already reported by Turner et al.~(2008). Our results reported in Table 3 are consistent overall with theirs.

\subsection{Mrk~509}

Observation 0130720101: the velocity width of the absorption line is $\sigma_v$$<$ 3800~km/s. From Fig.~12 in Appendix B, there is only one redshift solution. The initial baseline model statistic is $\chi^2$$=$151.0 for $d.o.f.$$=$153. Using an \emph{Xstar} grid with $v_t$$=$1000~km/s we obtain (1) log$\xi$$=$$5.20^{+0.17}_{-0.65}$~erg~s$^{-1}$~cm, $N_H$$>$$7 \times 10^{22}$~cm$^{-2}$ and $z$$=$$-0.132\pm0.003$, which corresponds to $v_{out}$$=$$0.173\pm0.003c$. The fit improvement is $\Delta \chi^2$$=$11.8 for three model parameters. Using an \emph{Xstar} grid with $v_t$$=$3000~km/s we obtain (2) log$\xi$$=$$5.55^{+0.09}_{-1.07}$~erg~s$^{-1}$~cm, $N_H$$>$$6 \times 10^{22}$~cm$^{-2}$ and $z$$=$$-0.132\pm0.004$, which corresponds to $v_{out}$$=$$0.173\pm0.004c$. The fit improvement is $\Delta \chi^2$$=$10.6 for three model parameters. The fits (1) and (2) are equivalent at 90\%. The best-fit is provided by (1), with F-test probability $P_F$$=$99.4\%. We detected a UFO.

Observation 0306090201: the velocity width of the absorption line is $\sigma_v$$<$ 3200~km/s. From Fig.~12 in Appendix B, there is only one redshift solution. The initial baseline model statistic is $\chi^2$$=$189.7 for $d.o.f.$$=$161. Using an \emph{Xstar} grid with $v_t$$=$1000~km/s we obtain (1) log$\xi$$=$$5.73^{+0.19}_{-1.69}$~erg~s$^{-1}$~cm, $N_H$$>$$6 \times 10^{22}$~cm$^{-2}$ and $z$$=$$-0.101\pm0.002$, which corresponds to $v_{out}$$=$$0.139\pm0.002c$. The fit improvement is $\Delta \chi^2$$=$12.7 for three model parameters. Using an \emph{Xstar} grid with $v_t$$=$3000~km/s we obtain (2) log$\xi$$=$$5.73^{+0.27}_{-1.17}$~erg~s$^{-1}$~cm, $N_H$$>$$10 \times 10^{22}$~cm$^{-2}$ and $z$$=$$-0.100\pm0.004$, which corresponds to $v_{out}$$=$$0.138\pm0.004c$. The fit improvement is $\Delta \chi^2$$=$10.1 for three model parameters. The best-fit is provided by (1), with F-test probability $P_F$$=$99\%. The fits (1) and (2) are equivalent at 90\%. We detected a UFO.

Observation 0306090401: the velocity width of the absorption line is $\sigma_v$$<$ 5000~km/s. From Fig.~12 in Appendix B, there are two possible redshift solutions, corresponding to an identification of the line with Fe~XXV He$\alpha$ or Fe~XXVI Ly$\alpha$. The initial baseline model statistic is $\chi^2$$=$215.5 for $d.o.f.$$=$162. Using an \emph{Xstar} grid with $v_t$$=$1000~km/s. For Fe~XXV we obtain (1) log$\xi$$=$$3.47^{+0.48}_{-0.15}$~erg~s$^{-1}$~cm, $N_H$$=$$(1.0^{+0.4}_{-0.3}) \times 10^{22}$~cm$^{-2}$ and $z$$=$$-0.185\pm0.003$, which corresponds to $v_{out}$$=$$0.234\pm0.003c$. The fit improvement is $\Delta \chi^2$$=$12.5 for three model parameters. Instead, for Fe~XXVI we obtain (2) log$\xi$$=$ $5.13^{+0.52}_{-0.73}$ erg~s$^{-1}$~cm, $N_H$$>$$7 \times 10^{22}$~cm$^{-2}$ and $z$$=$$-0.152\pm0.003$, which corresponds to $v_{out}$$=$$0.196\pm0.003c$. The fit improvement is $\Delta \chi^2$$=$14.5 for three model parameters. 
Using an \emph{Xstar} grid with $v_t$$=$3000~km/s. For Fe~XXV we obtain (3) log$\xi$$=$$2.97^{+0.25}_{-0.18}$~erg~s$^{-1}$~cm, $N_H$$=$$(0.8^{+0.3}_{-0.1}) \times 10^{22}$~cm$^{-2}$ and $z$$=$$-0.188\pm0.004$, which corresponds to $v_{out}$$=$$0.237\pm0.004c$. The fit improvement is $\Delta \chi^2$$=$12.2 for three model parameters. Instead, for Fe~XXVI we obtain (4) log$\xi$$=$ $5.54^{+0.17}_{-0.73}$ erg~s$^{-1}$~cm, $N_H$$>$$4 \times 10^{22}$~cm$^{-2}$ and $z$$=$$-0.153\pm0.004$, which corresponds to $v_{out}$$=$$0.197\pm0.004c$. The fit improvement is $\Delta \chi^2$$=$13.1 for three model parameters.
Using an \emph{Xstar} grid with $v_t$$=$5000~km/s. For Fe~XXV we obtain (5) log$\xi$$=$$3.48^{+0.54}_{-0.14}$~erg~s$^{-1}$~cm, $N_H$$=$$(1.0^{+0.3}_{-0.2}) \times 10^{22}$~cm$^{-2}$ and $z$$=$$-0.185\pm0.007$, which corresponds to $v_{out}$$=$$0.234\pm0.007c$. The fit improvement is $\Delta \chi^2$$=$10.1 for three model parameters. Instead, for Fe~XXVI we obtain (6) log$\xi$$=$$5.28^{+0.55}_{-0.90}$~erg~s$^{-1}$~cm, $N_H$$>$$4 \times 10^{22}$~cm$^{-2}$ and $z$$=$$-0.154\pm0.005$, which corresponds to $v_{out}$$=$$0.198\pm0.005c$. The fit improvement is $\Delta \chi^2$$=$11.5 for three model parameters. The best-fit is provided by (1), with F-test probability $P_F$$=$99\%. The fits (1), (2), (3) and (4) are equivalent at 90\%. The detection of UFOs in these observations was already reported by Cappi et al.~(2009). Our results reported in Table 3 are consistent overall with theirs.

\subsection{Ark~120}

Observation 0147190101: the velocity width of the absorption lines is $\sigma_v$$<$ 3300~km/s. From Fig.~12 in Appendix B, there are two possible redshift solutions, corresponding to an identification of the line with Fe~XXV He$\alpha$ or Fe~XXVI Ly$\alpha$. The initial baseline model statistic is $\chi^2$$=$161.8 for $d.o.f.$$=$156. Using an \emph{Xstar} grid with $v_t$$=$1000~km/s. For Fe~XXV we obtain (1) log$\xi$$=$$3.44^{+0.55}_{-0.18}$~erg~s$^{-1}$~cm, $N_H$$=$$(1.1^{+0.5}_{-0.4}) \times 10^{22}$~cm$^{-2}$ and $z$$=$$-0.247\pm0.003$, which corresponds to $v_{out}$$=$$0.306\pm0.003c$. The fit improvement is $\Delta \chi^2$$=$12.5 for three model parameters. Instead, for Fe~XXVI we obtain (2) log$\xi$$=$$5.27^{+0.58}_{-0.87}$~erg~s$^{-1}$~cm, $N_H$$>$$3 \times 10^{22}$~cm$^{-2}$ and $z$$=$$-0.216\pm0.003$, which corresponds to $v_{out}$$=$$0.269\pm0.003c$. The fit improvement is $\Delta \chi^2$$=$11.8 for three model parameters. 
Using an \emph{Xstar} grid with $v_t$$=$3000~km/s. For Fe~XXV we obtain (3) log$\xi$$=$$4.05^{+0.24}_{-1.33}$~erg~s$^{-1}$~cm, $N_H$$=$$(1.6^{+2.5}_{-1.1}) \times 10^{22}$~cm$^{-2}$ and $z$$=$$-0.247\pm0.003$, which corresponds to $v_{out}$$=$$0.306\pm0.003c$. The fit improvement is $\Delta \chi^2$$=$9.7 for three model parameters. Instead, for Fe~XXVI we obtain (4) log$\xi$$=$$5.68^{+0.07}_{-1.21}$~erg~s$^{-1}$~cm, $N_H$$>$$4 \times 10^{22}$~cm$^{-2}$ and $z$$=$$-0.216\pm0.004$, which corresponds to $v_{out}$$=$$0.269\pm0.004c$. The fit improvement is $\Delta \chi^2$$=$10.1 for three model parameters. The best-fit is provided by (1), with F-test probability $P_F$$=$99.4\%. The fits (1), (2) and (4) are equivalent at 90\%. We detected for the first time a UFO in this observation.

\subsection{Mrk~279}

Observation 0302480501: the velocity width of the absorption line is $\sigma_v$$<$ 4500~km/s. From Fig.~12 in Appendix B, there is only one redshift solution. The initial baseline model statistic is $\chi^2$$=$146.6 for $d.o.f.$$=$155. Using an \emph{Xstar} grid with $v_t$$=$1000~km/s we obtain (1) log$\xi$$=$$4.04^{+0.09}_{-0.53}$ erg~s$^{-1}$~cm, $N_H$$=$$(5.8^{+2.3}_{-3.3}) \times 10^{22}$~cm$^{-2}$ and $z$$=$$0.031\pm0.004$, which corresponds to $v_{out}$$<$$0.007c$. The fit improvement is $\Delta \chi^2$$=$14.0 for three model parameters. Using an \emph{Xstar} grid with $v_t$$=$3000~km/s we obtain (2) log$\xi$$=$$4.08^{+0.24}_{-0.95}$~erg~s$^{-1}$~cm, $N_H$$=$$ (3.7^{+1.5}_{-2.7}) \times 10^{22}$~cm$^{-2}$ and $z$$=$$0.031\pm0.005$, which corresponds to $v_{out}$$<$$0.009c$. The fit improvement is $\Delta \chi^2$$=$11.6 for three model parameters.  The best-fit is provided by (1), with F-test probability $P_F$$=$99.85\%. The fits (1) and (2) are equivalent at 90\%.

\subsection{Mrk~79}

Observation 0400070201: the velocity width of the absorption line is $\sigma_v$$<$ 6400~km/s. From Fig.~12 in Appendix B, there is only one redshift solution. The initial baseline model statistic is $\chi^2$$=$151.0 for $d.o.f.$$=$141. Using an \emph{Xstar} grid with $v_t$$=$1000~km/s we obtain (1) log$\xi$$=$$4.17^{+0.17}_{-0.21}$~erg~s$^{-1}$~cm, $N_H$$=$$(15.2^{+16.2}_{-7.9}) \times 10^{22}$~cm$^{-2}$ and $z$$=$$-0.067\pm0.003$, which corresponds to $v_{out}$$=$$0.091\pm0.003c$. The fit improvement is $\Delta \chi^2$$=$19.1 for three model parameters. Using an \emph{Xstar} grid with $v_t$$=$3000~km/s we obtain (2) log$\xi$$=$$4.34^{+0.08}_{-0.10}$~erg~s$^{-1}$~cm, $N_H$$=$$ (18.1^{+4.8}_{-8.1}) \times 10^{22}$~cm$^{-2}$ and $z$$=$$-0.068\pm0.003$, which corresponds to $v_{out}$$=$$0.092\pm0.003c$. $\Delta \chi^2$$=$16.5 for three model parameters. 
Using an \emph{Xstar} grid with $v_t$$=$5000~km/s we obtain (3) log$\xi$$=$$4.29^{+0.10}_{-0.17}$~erg~s$^{-1}$~cm, $N_H$$=$$ (11.5^{+5.4}_{-5.9}) \times 10^{22}$~cm$^{-2}$ and $z$$=$$-0.068\pm0.005$, which corresponds to $v_{out}$$=$$0.091\pm0.005c$. The fit improvement is $\Delta \chi^2$$=$13.5 for three model parameters. The best-fit is provided by (1), with F-test probability $P_F$$=$99.97\%. The fits (1) and (2) are equivalent at 90\%. We detected for the first time a UFO in this observation.

\subsection{NGC~4051}

Observation 0109141401: the velocity width of the absorption line is $\sigma_v$$<$ 4200~km/s. From Fig.~12 in Appendix B, there are two possible redshift solutions, corresponding to an identification of the line with Fe~XXV He$\alpha$ or Fe~XXVI Ly$\alpha$. The initial baseline model statistic is $\chi^2$$=$174.3 for $d.o.f.$$=$158. Using an \emph{Xstar} grid with $v_t$$=$1000~km/s. For Fe~XXV we obtain (1) log$\xi$$=$$3.64^{+0.45}_{-0.18}$~erg~s$^{-1}$~cm, $N_H$$=$$(1.1^{+1.5}_{-0.3}) \times 10^{22}$~cm$^{-2}$ and $z$$=$$-0.051\pm0.003$, which corresponds to $v_{out}$$=$$0.054\pm0.003c$. The fit improvement is $\Delta \chi^2$$=$13.8 for three model parameters. Instead, for Fe~XXVI we obtain (2) log$\xi$$=$$5.73^{+0.18}_{-1.06}$~erg~s$^{-1}$~cm, $N_H$$>$$6 \times 10^{22}$~cm$^{-2}$ and $z$$=$$-0.014\pm0.005$, which corresponds to $v_{out}$$=$$0.016\pm0.005c$. The fit improvement is $\Delta \chi^2$$=$11.8 for three model parameters. 
Using an \emph{Xstar} grid with $v_t$$=$3000~km/s. For Fe~XXV we obtain (3) log$\xi$$=$$4.18^{+1.20}_{-1.36}$~erg~s$^{-1}$~cm, $N_H$$=$$(2.4^{+1.6}_{-1.5}) \times 10^{22}$~cm$^{-2}$ and $z$$=$$-0.054\pm0.005$, which corresponds to $v_{out}$$=$$0.057\pm0.005c$. The fit improvement is $\Delta \chi^2$$=$12.0 for three model parameters. Instead, for Fe~XXVI we obtain (4) log$\xi$$=$ $5.69^{+0.06}_{-1.21}$ erg~s$^{-1}$~cm, $N_H$$>$$5 \times 10^{22}$~cm$^{-2}$ and $z$$=$$-0.015\pm0.005$, which corresponds to $v_{out}$$=$$0.017\pm0.005c$. The fit improvement is $\Delta \chi^2$$=$11.5 for three model parameters. The best-fit is provided by (1), with F-test probability $P_F$$=$99.5\%. The fits (1), (2), (3) and (4) are equivalent at 90\%. Considering the mean value of the outflow velocity between the two line identifications in Table 3, this absorber is consistent with a UFO.

Observation 0157560101: velocity width absorption line is measured as $\sigma_v$$=$ $4700^{+1800}_{-1500}$~km/s. We used an \emph{Xstar} grid with $v_t$$=$5000~km/s. From Fig.~12 in Appendix B there is only one preferred solution. We obtain log$\xi$$=$$3.06^{+0.17}_{-0.19}$~erg~s$^{-1}$~cm, $N_H$$=$$(5.2^{+1.1}_{-1.3})\times 10^{22}$~cm$^{-2}$ and $z$$=$$-0.183\pm0.006$, which corresponds to $v_{out}$$=$$0.202\pm0.006c$. The initial baseline model statistic is $\chi^2$$=$174.6 for $d.o.f.$$=$134. The fit improvement is $\Delta \chi^2$$=$22.0 for three model parameters, which corresponds to an F-test probability of $P_F$$=$99.95\%. We detected for the first time a UFO in this observation.

\subsection{Mrk~766}

Observation 0304030301: the velocity width of the absorption line is measured as $\sigma_v$$=$ $4600^{+2700}_{-1800}$~km/s. We used an \emph{Xstar} grid with $v_t$$=$5000~km/s. From Fig.~12 in Appendix B there is only one solution. We obtain log$\xi$$=$$3.46^{+0.71}_{-0.32}$ erg~s$^{-1}$~cm, $N_H$$=$$(2.1^{+0.4}_{-0.3})\times 10^{22}$~cm$^{-2}$ and $z$$=$$-0.067\pm0.006$, which corresponds to $v_{out}$$=$$0.082\pm0.006c$. The initial baseline model statistic is $\chi^2$$=$205.0 for $d.o.f.$$=$157. The fit improvement is $\Delta \chi^2$$=$22.4 for three model parameters, which corresponds to an F-test probability of $P_F$$=$99.95\%. We detected a UFO.

Observation 0304030501: the velocity width of the absorption line is $\sigma_v$$<$ 4400~km/s. From Fig.~12 in Appendix B, there is only one redshift solution. The initial baseline model statistic is $\chi^2$$=$208.3 for $d.o.f.$$=$154. Using an \emph{Xstar} grid with $v_t$$=$1000~km/s we obtain (1) log$\xi$$=$$4.28^{+0.08}_{-0.11}$~erg~s$^{-1}$~cm, $N_H$$=$$(10.0^{+6.4}_{-3.5}) \times 10^{22}$~cm$^{-2}$ and $z$$=$$-0.072\pm0.002$, which corresponds to $v_{out}$$=$$0.088\pm0.002c$. The fit improvement is $\Delta \chi^2$$=$21.2 for three model parameters. Using an \emph{Xstar} grid with $v_t$$=$3000~km/s we obtain (2) log$\xi$$=$$4.40^{+0.07}_{-0.06}$~erg~s$^{-1}$~cm, $N_H$$=$$(9.2^{+4.6}_{-2.8}) \times 10^{22}$~cm$^{-2}$ and $z$$=$$-0.074\pm0.003$, which corresponds to $v_{out}$$=$$0.090\pm0.003c$. The fit improvement is $\Delta \chi^2$$=$17.3 for three model parameters. The best-fit is provided by (1), with F-test probability $P_F$$=$99.9\%. The fit (1) is better than (2) at 90\% level. We detected for the first time a UFO in this observation. The detection of a possible UFO in the first observation was already suggested by Miller et al.~(2007) and Turner et al.~(2007). Our result reported in Table 3 is consistent overall with theirs.

\subsection{Mrk~841}

Observation 0205340401: the velocity width of the absorption line is $\sigma_v$$<$ 4000~km/s. From Fig.~12 in Appendix B, there are two possible redshift solutions, corresponding to an identification of the line with Fe~XXV He$\alpha$ or Fe~XXVI Ly$\alpha$. The initial baseline model statistic is $\chi^2$$=$147.6 for $d.o.f.$$=$130. Using an \emph{Xstar} grid with $v_t$$=$1000~km/s. For Fe~XXV we obtain (1) log$\xi$$=$$3.46^{+0.38}_{-0.24}$~erg~s$^{-1}$~cm, $N_H$$=$$(3.5^{+1.9}_{-1.3}) \times 10^{22}$~cm$^{-2}$ and $z$$=$$-0.035\pm0.004$, which corresponds to $v_{out}$$=$$0.071\pm0.004c$. The fit improvement is $\Delta \chi^2$$=$12.2 for three model parameters. Instead, for Fe~XXVI we obtain (2) log$\xi$$=$$4.50^{+0.65}_{-0.24}$~erg~s$^{-1}$~cm, $N_H$$>$$11 \times 10^{22}$~cm$^{-2}$ and $z$$=$$0.003\pm0.004$, which corresponds to $v_{out}$$=$$0.034\pm0.004c$. The fit improvement is $\Delta \chi^2$$=$12.7 for three model parameters. 
Using an \emph{Xstar} grid with $v_t$$=$3000~km/s. For Fe~XXV we obtain (3) log$\xi$$=$$2.93^{+1.03}_{-0.27}$~erg~s$^{-1}$~cm, $N_H$$=$$(2.3^{+3.5}_{-1.3}) \times 10^{22}$~cm$^{-2}$ and $z$$=$$-0.038\pm0.005$, which corresponds to $v_{out}$$=$$0.075\pm0.005c$. The fit improvement is $\Delta \chi^2$$=$11.9 for three model parameters. Instead, for Fe~XXVI we obtain (4) log$\xi$$=$$4.54^{+0.09}_{-0.08}$~erg~s$^{-1}$~cm, $N_H$$>$$10 \times 10^{22}$~cm$^{-2}$ and $z$$=$$0.003\pm0.005$, which corresponds to $v_{out}$$=$$0.034\pm0.005c$. The fit improvement is $\Delta \chi^2$$=$12.4 for three model parameters. The best-fit is provided by (2), with F-test probability $P_F$$=$99.07\%. The fits (1), (2), (3) and (4) are equivalent at 90\%. We detected for the first time a UFO in this observation.

\subsection{ESO~323-G77}

Observation 0300240501: the velocity width of the absorption lines is $\sigma_v$$<$ 3700~km/s. From Fig.~12 in Appendix B, there is only one redshift solution. The initial baseline model statistic is $\chi^2$$=$223.8 for $d.o.f.$$=$135. Using an \emph{Xstar} grid with $v_t$$=$1000~km/s we obtain (1) log$\xi$$=$$3.45^{+0.07}_{-0.09}$~erg~s$^{-1}$~cm, $N_H$$=$$(13.3^{+3.8}_{-2.2}) \times 10^{22}$~cm$^{-2}$ and $z$$=$$0.004\pm0.003$, which corresponds to $v_{out}$$=$$0.011\pm0.003c$. The fit improvement is $\Delta \chi^2$$=$74.3 for three model parameters. Using an \emph{Xstar} grid with $v_t$$=$3000~km/s we obtain (2) log$\xi$$=$$4.03^{+0.08}_{-0.06}$~erg~s$^{-1}$~cm, $N_H$$=$$(29.1^{+6.9}_{-6.2}) \times 10^{22}$~cm$^{-2}$ and $z$$=$$0.003\pm0.003$, which corresponds to $v_{out}$$=$$0.012\pm0.003c$. The fit improvement is $\Delta \chi^2$$=$77.0 for three model parameters. The best-fit is provided by (2), with F-test probability $P_F$$>$99.99\%. The fits (1) and (2) are equivalent at 90\%. The detection of an Fe K absorber was already reported by Jim{\'e}nez-Bail{\'o}n et al.~(2008). Our results reported in Table 3 are consistent overall with theirs.

\subsection{1H0419-577}

Observation 0148000201: the velocity width of the absorption line is $\sigma_v$$<$ 7000~km/s. From Fig.~12 in Appendix B, there is only one redshift solution. The initial baseline model statistic is $\chi^2$$=$128.7 for $d.o.f.$$=$105. Using an \emph{Xstar} grid with $v_t$$=$1000~km/s we obtain (1) log$\xi$$=$$3.85^{+0.18}_{-0.38}$~erg~s$^{-1}$~cm, $N_H$$=$$(17.2^{+14.3}_{-8.3}) \times 10^{22}$~cm$^{-2}$ and $z$$=$$0.023\pm0.004$, which corresponds to $v_{out}$$=$$0.076\pm0.004c$. The fit improvement is $\Delta \chi^2$$=$14.8 for three model parameters. Using an \emph{Xstar} grid with $v_t$$=$3000~km/s we obtain (2) log$\xi$$=$$3.89^{+0.33}_{-0.73}$~erg~s$^{-1}$~cm, $N_H$$=$$(11.3^{+10.7}_{-8.3}) \times 10^{22}$~cm$^{-2}$ and $z$$=$$0.019\pm0.005$, which corresponds to $v_{out}$$=$$0.080\pm0.005c$. The fit improvement is $\Delta \chi^2$$=$12.2 for three model parameters. Using an \emph{Xstar} grid with $v_t$$=$5000~km/s we obtain (3) log$\xi$$=$$3.70^{+0.47}_{-0.55}$~erg~s$^{-1}$~cm, $N_H$$=$$(4.2^{+7.8}_{-1.5}) \times 10^{22}$~cm$^{-2}$ and $z$$=$$0.021\pm0.006$, which corresponds to $v_{out}$$=$$0.078\pm0.006c$. The fit improvement is $\Delta \chi^2$$=$11.1 for three model parameters. The best-fit is provided by (1), with F-test probability $P_F$$>$99.43\%. The fits (1) and (2) are equivalent at 90\%. We detected for the first time a UFO in this observation.

\subsection{Mrk~290}

Observation 0400360601: the velocity width of the absorption line is $\sigma_v$$<$ 4000~km/s. From Fig.~12 in Appendix B, there are two possible redshift solutions, corresponding to an identification of the line with Fe~XXV He$\alpha$ or Fe~XXVI Ly$\alpha$. The initial baseline model statistic is $\chi^2$$=$103.7 for $d.o.f.$$=$104. 
Using an \emph{Xstar} grid with $v_t$$=$1000~km/s we obtain (1) for Fe~XXV log$\xi$$=$$3.54^{+0.20}_{-0.25}$~erg~s$^{-1}$~cm, $N_H$$=$$(4.7^{+4.4}_{-1.6}) \times 10^{22}$~cm$^{-2}$ and $z$$=$$-0.142\pm0.004$, which corresponds to $v_{out}$$=$$0.181\pm0.004c$. The fit improvement is $\Delta \chi^2$$=$11.6 for three model parameters. Instead, for Fe~XXVI we obtain (2) log$\xi$$=$$4.37^{+0.42}_{-0.25}$~erg~s$^{-1}$~cm, $N_H$$>$$11.5 \times 10^{22}$~cm$^{-2}$ and $z$$=$$-0.107\pm0.004$, which corresponds to $v_{out}$$=$$0.142\pm0.004c$. The fit improvement is $\Delta \chi^2$$=$13.3 for three model parameters. 
Using an \emph{Xstar} grid with $v_t$$=$3000~km/s we obtain (3) for Fe~XXV log$\xi$$=$$3.07^{+0.56}_{-0.33}$~erg~s$^{-1}$~cm, $N_H$$=$$(3.1^{+1.2}_{-1.2}) \times 10^{22}$~cm$^{-2}$ and $z$$=$$-0.143\pm0.005$, which corresponds to $v_{out}$$=$$0.182\pm0.005c$. The fit improvement is $\Delta \chi^2$$=$12.0 for three model parameters. Instead for Fe~XXVI we obtain (4) log$\xi$$=$$4.35^{+0.73}_{-0.13}$~erg~s$^{-1}$~cm, $N_H$$=$$(33.3^{+16.6}_{-11.1}) \times 10^{22}$~cm$^{-2}$ and $z$$=$$-0.107\pm0.004$, which corresponds to $v_{out}$$=$$0.142\pm0.004c$. The fit improvement is $\Delta \chi^2$$=$13.5 for three model parameters. The best-fit is provided by (4), with F-test probability $P_F$$>$99.74\%. The fits (1), (2), (3) and (4) are equivalent at 90\%. We detected for the first time a UFO in this observation.

\subsection{Mrk~205}

Observation 0124110101: the velocity width of the absorption line is $\sigma_v$$<$ 3100~km/s. From Fig.~12 in Appendix B, there is only one redshift solution. The initial baseline model statistic is $\chi^2$$=$135.7 for $d.o.f.$$=$121. Using an \emph{Xstar} grid with $v_t$$=$1000~km/s we obtain (1) log$\xi$$=$$4.86^{+0.24}_{-0.90}$~erg~s$^{-1}$~cm, $N_H$$>$$14.5 \times 10^{22}$~cm$^{-2}$ and $z$$=$$-0.031\pm0.004$, which corresponds to $v_{out}$$=$$0.100\pm0.004c$. The fit improvement is $\Delta \chi^2$$=$11.9 for three model parameters. Using an \emph{Xstar} grid with $v_t$$=$3000~km/s we obtain (2) log$\xi$$=$$5.31^{+0.16}_{-0.74}$~erg~s$^{-1}$~cm, $N_H$$>$$22.7 \times 10^{22}$~cm$^{-2}$ and $z$$=$$-0.031\pm0.004$, which corresponds to $v_{out}$$=$$0.100\pm0.004$. The fit improvement is $\Delta \chi^2$$=$10.1 for three model parameters. The fits (1) and (2) are equivalent at 90\%. The best-fit is provided by (1), with F-test probability $P_F$$>$98.8\%. We detected for the first time a UFO in this observation.

\subsection{PG~1211+143}

Observation 0112610101: the velocity width of the absorption line is measured as $\sigma_v$$=$ $4500^{+1500}_{-1200}$~km/s. We used an \emph{Xstar} grid with $v_t$$=$5000~km/s. From Fig.~12 in Appendix B, there is only one solution. We obtained log$\xi$$=$$2.87^{+0.12}_{-0.10}$ erg~s$^{-1}$~cm, $N_H$$=$$(8.0^{+2.2}_{-1.1})\times 10^{22}$~cm$^{-2}$ and $z$$=$$-0.071\pm0.003$, which corresponds to $v_{out}$$=$$0.151\pm0.003c$. The initial baseline model statistic is $\chi^2$$=$196.0 for $d.o.f.$$=$117. The fit improvement is $\Delta \chi^2$$=$70.5 for three model parameters, which corresponds to an F-test probability of $P_F$$>$99.99\%. We detected a UFO. The detection of a UFO in this observation was already reported by Pounds et al.~(2003) and Pounds \& Page (2006). Our results reported in Table 3 are consistent overall with theirs.

\subsection{MCG-5-23-16}

Observation 0302850201: the velocity width of the absorption line is $\sigma_v$$<$ 13000~km/s. From Fig.~12 in Appendix B, there is only one redshift solution. The initial baseline model statistic is $\chi^2$$=$298.4 for $d.o.f.$$=$161. Using an \emph{Xstar} grid with $v_t$$=$1000~km/s we obtain (1) log$\xi$$=$$4.30^{+0.06}_{-0.10}$~erg~s$^{-1}$~cm, $N_H$$=$$(3.2^{+0.8}_{-0.9}) \times 10^{22}$~cm$^{-2}$ and $z$$=$$-0.101\pm0.003$, which corresponds to $v_{out}$$=$$0.115\pm0.003c$. The fit improvement is $\Delta \chi^2$$=$34.0 for three model parameters. Using an \emph{Xstar} grid with $v_t$$=$3000~km/s we obtain (2) log$\xi$$=$$4.38^{+0.03}_{-0.03}$~erg~s$^{-1}$~cm, $N_H$$=$$(4.3^{+0.6}_{-0.5}) \times 10^{22}$~cm$^{-2}$ and $z$$=$$-0.103\pm0.003$ which corresponds to $v_{out}$$=$$0.117\pm0.003c$. The fit improvement is $\Delta \chi^2$$=$41.3 for three model parameters. 
Using an \emph{Xstar} grid with $v_t$$=$5000~km/s we obtain (3) log$\xi$$=$$4.31^{+0.06}_{-0.06}$~erg~s$^{-1}$~cm, $N_H$$=$$(3.8^{+1.3}_{-1.0}) \times 10^{22}$~cm$^{-2}$ and $z$$=$$-0.102\pm0.004$, which corresponds to $v_{out}$$=$$0.116\pm0.004c$. The fit improvement is $\Delta \chi^2$$=$41.0 for three model parameters. The best-fit is provided by (2), with F-test probability $P_F$$>$99.99\%. The fits (2) and (3) are equivalent at 90\%. The detection of a UFO in this observation was already suggested by Braito et al.~(2007). Our results reported in Table 3 are consistent overall with theirs.

\subsection{NGC~4507}

Observation 0006220201: the velocity width of the absorption line is $\sigma_v$$<$ 4200~km/s. From Fig.~12 in Appendix B, there are two possible redshift solutions, corresponding to an identification of the line with Fe~XXV He$\alpha$ or Fe~XXVI Ly$\alpha$. Using an \emph{Xstar} grid with $v_t$$=$1000~km/s we obtain (1) for Fe~XXV log$\xi$$=$$3.81^{+0.31}_{-0.43}$~erg~s$^{-1}$~cm, $N_H$$=$$(2.3^{+2.8}_{-1.3}) \times 10^{22}$~cm$^{-2}$ and $z$$=$$-0.188\pm0.003$, which corresponds to $v_{out}$$=$$0.216\pm0.003c$. The fit improvement is $\Delta \chi^2$$=$10.1 for three model parameters. Instead, for Fe~XXVI we obtain (2) log$\xi$$=$$5.36^{+0.32}_{-0.93}$~erg~s$^{-1}$~cm, $N_H$$>$$10 \times 10^{22}$~cm$^{-2}$ and $z$$=$$-0.155\pm0.003$, which corresponds to $v_{out}$$=$$0.178\pm0.003c$. The fit improvement is $\Delta \chi^2$$=$8.7 for three model parameters.
Using an \emph{Xstar} grid with $v_t$$=$3000~km/s we obtain (3) for Fe~XXV log$\xi$$=$$4.16^{+0.14}_{-0.79}$~erg~s$^{-1}$~cm, $N_H$$=$$(4.1^{+2.8}_{-3.2}) \times 10^{22}$~cm$^{-2}$ and $z$$=$$-0.190\pm0.004$, which corresponds to $v_{out}$$=$$0.218\pm0.004c$. The fit improvement is $\Delta \chi^2$$=$9.7 for three model parameters. Instead, for Fe~XXVI we obtain (4) log$\xi$$=$$4.78^{+0.89}_{-0.37}$~erg~s$^{-1}$~cm, $N_H$$>$$5 \times 10^{22}$~cm$^{-2}$ and $z$$=$$-0.156\pm0.004$, which corresponds to $v_{out}$$=$$0.179\pm0.004c$. The fit improvement is $\Delta \chi^2$$=$8.3 for three model parameters. The best-fit is provided by (1), with F-test probability $P_F$$=$97.7\%. The fits (1), (2), (3) and (4) are equivalent at 90\%. We detected for the first time a UFO in this observation.

\subsection{NGC~7582}

Observation 0112310201: the velocity width of the absorption line is $\sigma_v$$<$ 11000~km/s. From Fig.~12 in Appendix B, there is only one redshift solution. The initial baseline model statistic is $\chi^2$$=$131.9 for $d.o.f.$$=$103. Using an \emph{Xstar} grid with $v_t$$=$1000~km/s we obtain (1) log$\xi$$=$$3.39^{+0.09}_{-0.15}$~erg~s$^{-1}$~cm, $N_H$$=$$(23.4^{+21.1}_{-9.8}) \times 10^{22}$~cm$^{-2}$ and $z$$=$$-0.250\pm0.003$, which corresponds to $v_{out}$$=$$0.285\pm0.003c$. The fit improvement is $\Delta \chi^2$$=$20.9 for three model parameters. Using an \emph{Xstar} grid with $v_t$$=$3000~km/s we obtain (2) log$\xi$$=$$3.74^{+0.27}_{-0.81}$~erg~s$^{-1}$~cm, $N_H$$=$$(11.0^{+16.2}_{-6.9}) \times 10^{22}$~cm$^{-2}$ and $z$$=$$-0.250\pm0.003$, which corresponds to $v_{out}$$=$$0.285\pm0.003c$. The fit improvement is $\Delta \chi^2$$=$18.1 for three model parameters. Using an \emph{Xstar} grid with $v_t$$=$5000~km/s we obtain (3) log$\xi$$=$$3.31^{+0.19}_{-0.25}$~erg~s$^{-1}$~cm, $N_H$$=$$(6.8^{+2.4}_{-1.8}) \times 10^{22}$~cm$^{-2}$ and $z$$=$$-0.248\pm0.005$, which corresponds to $v_{out}$$=$$0.283\pm0.005c$. The fit improvement is $\Delta \chi^2$$=$15.6 for three model parameters. The best-fit is provided by (1), with F-test probability $P_F$$=$99.94\%. The fit (1) is better than (2) and (3) at the 90\% level. We detected for the first time a UFO in this observation.

We note that TA reported the simultaneous evidence for an absorption line at E$\simeq$9~keV, possibly due to Fe XXV, and three additional weak lines in the energy range E$\simeq$4--5.3~keV, possibly due to He/H-like Ar and Ca, in this observation.
The significance of the lower energy lines is only marginal, being in the range $\sim$95--99\%, and their modeling can be hampered by the intrinsic complexity of the spectrum of this source at energies lower than 6~keV, showing the superimposition of different emission/absorption components (Piconcelli et al.~2007). 
However, as already discussed in TA, if the centroids of the lower energy lines are fixed to the expected energies for Ar~XVII He$\alpha$ (E$\simeq$3.1~keV), Ar~XVII Ly$\alpha$ (E$\simeq$3.3~keV) and Ca~XIX He$\alpha$ (E$\simeq$3.9~keV) (Verner et al.~1996) and they are fitted together with the Fe~XXV He$\alpha$ line, it turns out that they are consistent with sharing the same common blue-shift of $z=$$-0.255\pm0.002$. This modeling provides a good representation of the data, with a fit improvement of $\Delta\chi^2$$=$35 for five parameters, which corresponds to a very low random probability of less than $10^{-5}$. This was already discussed in \S4.5 and Appendix B of TA. However, from the ratios of the EWs of the lower energy lines with respect to the Fe~XXV He$\alpha$ we can also roughly estimate that there is the need for a $\sim$5 overabundance of lighter elements, such as Ar and Ca, with respect to Fe. In particular, even though in this case the statistics is presumably mostly driven by the intense Fe~XXV/XXVI absorption lines, a single \emph{Xstar} grid seems capable to simultaneously parametrize all the lines, but those at lower energy are more intense than what expected assuming standard Solar abundances. An \emph{Xstar} grid with $\sim$5 overabundance of Ar and Ca actually provides a better modeling of these lines. However, the detailed analysis of possible low energy lines is beyond the scope of the present paper, which is focused on the study of the UFOs essentially through the more intense Fe K-shell lines. We postpone their treatment to forthcoming works.

\clearpage

\section{$\chi^2$ statistic and observed absorber redshift contour plots}

   \begin{figure}[!hb]
   \centering
    \includegraphics[width=12cm,height=15cm,angle=0]{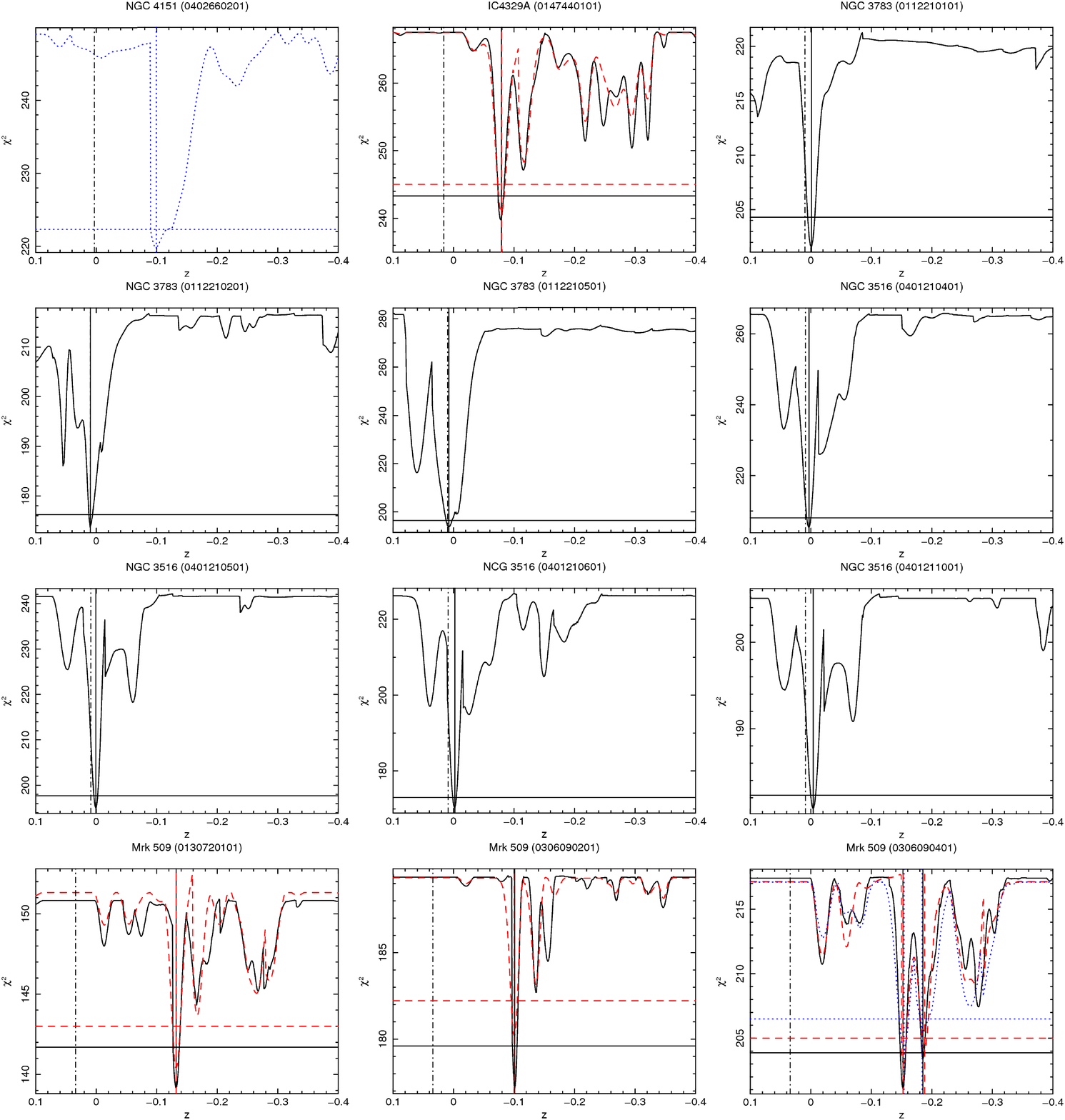}
   \caption{{\small $\chi^2$ statistic plotted against the \emph{Xstar} absorber redshift. The black solid, dashed red and dotted blue contours refer to \emph{Xstar} grids with turbulent velocities of 1000~km/s, 3000~km/s and 5000~km/s, respectively. Not all observations required grids with different turbulent velocities, see text for more details. The horizontal lines indicate the 90\% confidence levels for one interesting parameter. The vertical dash-dotted line indicates the cosmological redshift of the source. The remaining vertical lines indicate the absorber redshift from the \emph{Xstar} fits. In this case a negative redshift indicates a blueshift.}}
    \end{figure}

   \begin{figure}[htb]
   \centering
   \includegraphics[width=12cm,height=15cm,angle=0]{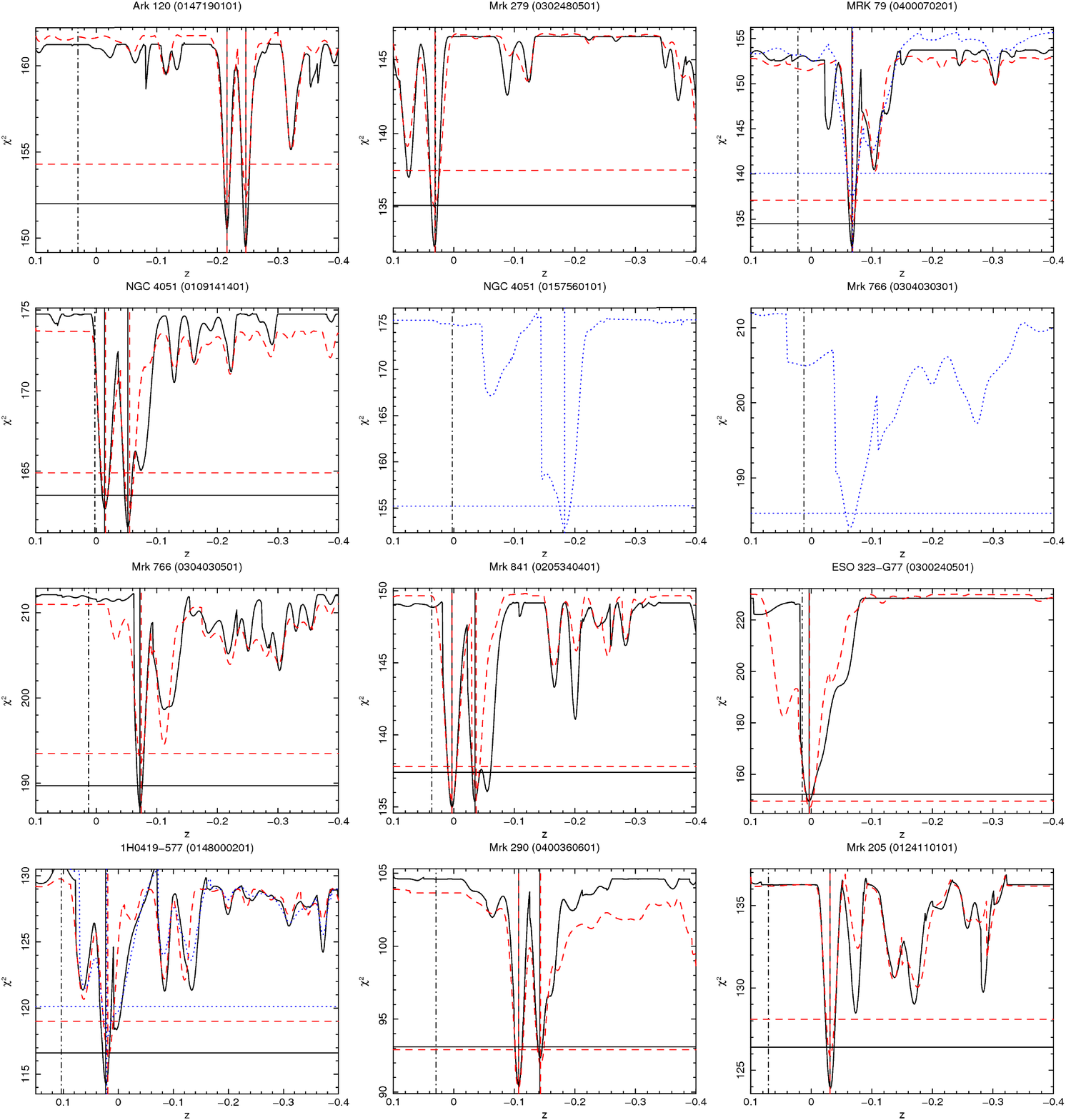}
   \caption{{\small  Continued.}}
    \end{figure}

   \begin{figure}[htb]
   \centering
    \includegraphics[width=9.5cm,height=8.5cm,angle=0]{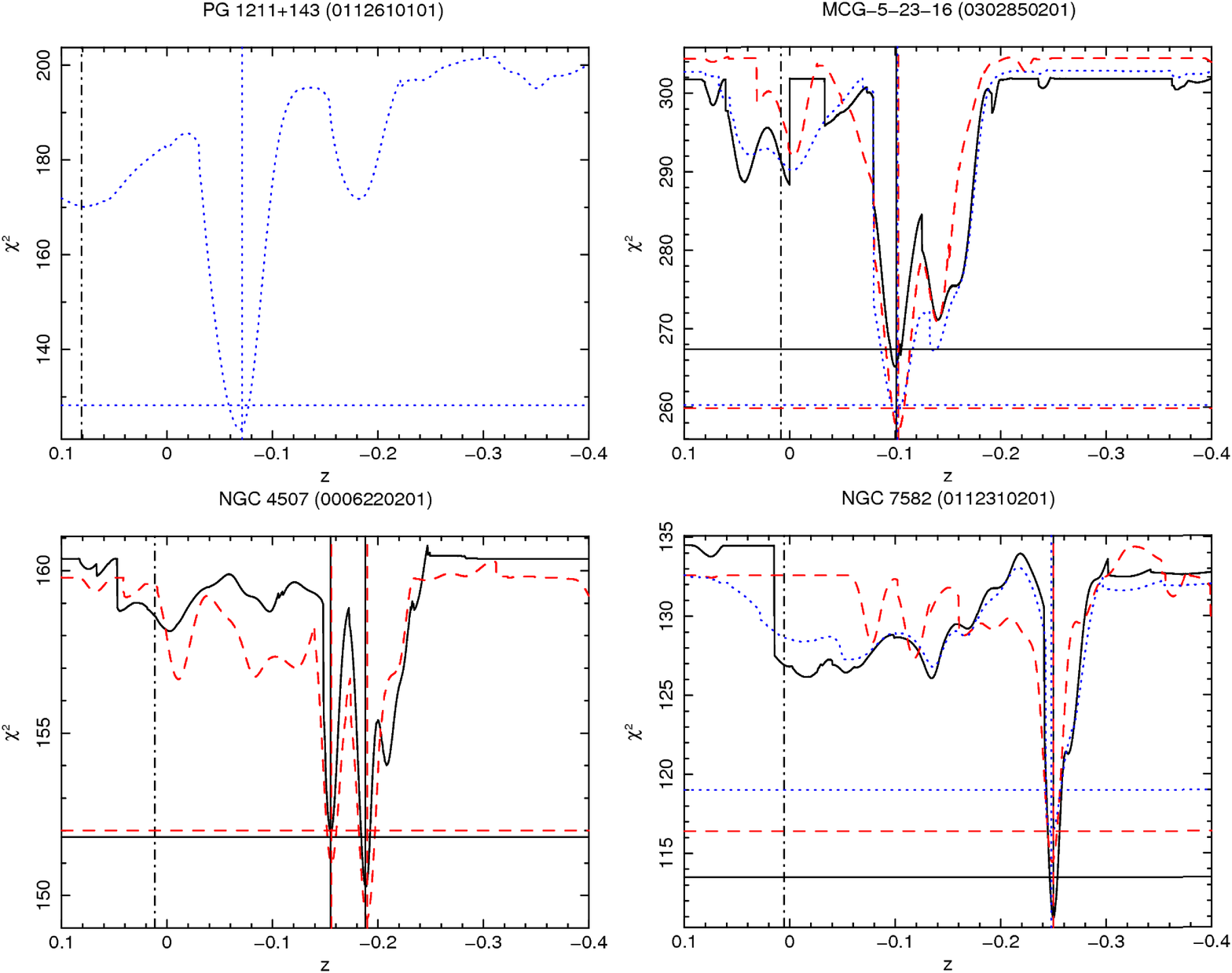}
   \caption{{\small  Continued.}}
    \end{figure}


\begin{thebibliography}{}


\bibitem[]{} Anders, E., \& Grevesse, N.\ 1989, \gca, 53, 197
\bibitem{1993ARA&A..31..473A} Antonucci, R.\ 1993, ARA\&A, 31, 473 
\bibitem[Aslplund et al.(2009)]{2009ARA&A..47..481A} Asplund, M., Grevesse, N., Sauval, A.~J., \& Scott, P.\ 2009, \araa, 47, 481
\bibitem[Bianchi et al.(2005)]{2005MNRAS.357..599B} Bianchi, S., Matt, G., Nicastro, F., Porquet, D., \& Dubau, J.\ 2005, \mnras, 357, 599
\bibitem[blustin (2005)]{blustin05} Blustin, A.~J., Page, M.~J., Fuerst, S.~V., Branduardi-Raymont, G., \& Ashton, C.~E.\ 2005, \aap, 431, 111
\bibitem[Braito (2007)]{braito07} Braito, V., et al.\ 
2007, \apj, 670, 978
\bibitem[cappi (2006)]{cappi06} Cappi, M.\ 2006, Astronomische 
Nachrichten, 327, 1012
\bibitem[cappi (2009)]{cappi09} Cappi, M., et al.\ 2009, \aap, 504, 401
\bibitem[Cash(1979)]{1979ApJ...228..939C} Cash, W.\ 1979, \apj, 228, 939
\bibitem[Chartas et al.(2002)]{2002ApJ...579..169C} Chartas, G., Brandt, W.~N., Gallagher, S.~C., \& Garmire, G.~P.\ 2002, \apj, 579, 169
\bibitem[Chartas et al.(2003)]{2003ApJ...595...85C} Chartas, G., Brandt, W.~N., \& Gallagher, S.~C.\ 2003, \apj, 595, 85
\bibitem[Crenshaw et al.(2003)]{2003ARA&A..41..117C} Crenshaw, D.~M., Kraemer, S.~B., \& George, I.~M.\ 2003, \araa, 41, 117 
\bibitem[dadina (2005)]{dadina05} Dadina, M., Cappi, M., Malaguti, G., Ponti, G., \& de Rosa, A.\ 2005, \aap, 442, 461
\bibitem[Dadina(2008)]{2008A&A...485..417D} Dadina, M.\ 2008, \aap, 485, 417
\bibitem{1994ApJS...95....1E} Elvis, M., et al.\ 1994, ApJS, 95, 1
\bibitem{2000ApJ...545...63E} Elvis, M.\ 2000, ApJ, 545, 63 
\bibitem[Elvis(2006)]{2006MmSAI..77..573E} Elvis, M.\ 2006, Memorie della Societa Astronomica Italiana, 77, 573
\bibitem[Fabian(2009)]{2009arXiv0912.0880F} Fabian, A.~C.\ 2009, arXiv:0912.0880
\bibitem[Ferland(2003)]{2003ARA&A..41..517F} Ferland, G.~J.\ 2003, \araa, 41, 517 
\bibitem[Ghisellini et al.(2004)]{2004A&A...413..535G} Ghisellini, G., Haardt, F., \& Matt, G.\ 2004, \aap, 413, 535
\bibitem[Guainazzi]{Guainazzi} Guainazzi, M., et al.~2010, XMM-SOC-CAL-TN-0018
\bibitem{Haardt 1991} Haardt, F., Maraschi, L., 1991, ApJ, 380, L51
\bibitem[jimenez (2008)]{jimenez08} Jim{\'e}nez-Bail{\'o}n, E., Krongold, Y., Bianchi, S., Matt, G., Santos-Lle{\'o}, M., Piconcelli, E., \& Schartel, N.\ 2008, \mnras, 391, 1359
\bibitem[kallman (2001)]{kallman01} Kallman, T., \& Bautista, M.\ 2001, \apjs, 133, 221
\bibitem[kallman (2004)]{kallman04} Kallman, T.~R., Palmeri, P., Bautista, M.~A., Mendoza, C., \& Krolik, J.~H.\ 2004, \apjs, 155, 675
\bibitem[katayama (2004)]{katayama04} Katayama, H., Takahashi, I., Ikebe, Y., Matsushita, K., \& Freyberg, M.~J.\ 2004, \aap, 414, 767
\bibitem[Kellermann et al.(1989)]{1989AJ.....98.1195K} Kellermann, K.~I., Sramek, R., Schmidt, M., Shaffer, D.~B., \& Green, R.\ 1989, \aj, 98, 1195
\bibitem[king (2003)]{king03} King, A.~R., \& Pounds, K.~A.\ 2003, \mnras, 345, 657
\bibitem[King(2010)]{2010MNRAS.402.1516K} King, A.~R.\ 2010, \mnras, 402, 1516 (King 2010a)
\bibitem[King(2010)]{2010MNRAS.408L..95K} King, A.~R.\ 2010, \mnras, 408, L95 (King 2010b)
\bibitem[Kinkhabwala et al.(2002)]{2002ApJ...575..732K} Kinkhabwala, A., et 
al.\ 2002, \apj, 575, 732 
\bibitem[markowitz (2006)]{mark06} Markowitz, A., 
Reeves, J.~N., \& Braito, V.\ 2006, \apj, 646, 783
\bibitem{2003ApJ...593..142M} McKernan, B., Yaqoob, T., George, I.~M., \& Turner, T.~J.\ 2003, ApJ, 593, 142 
\bibitem[mckernan (2007)]{mckernan07} McKernan, B., Yaqoob, 
T., \& Reynolds, C.~S.\ 2007, \mnras, 379, 1359
\bibitem[miller (2007)]{miller07} Miller, L., Turner, T.~J., Reeves, J.~N., George, I.~M., Kraemer, S.~B., \& Wingert, B.\ 2007, \aap, 463, 131
\bibitem[Nicastro et al.(1999)]{1999ApJ...517..108N} Nicastro, F., Fiore, F., \& Matt, G.\ 1999, \apj, 517, 108 
\bibitem[ohsuga (2009)]{ohsuga09} Ohsuga, K., Mineshige, 
S., Mori, M., \& Kato, Y.\ 2009, \pasj, 61, L7
\bibitem[palmeri (2002)]{palmeri02} Palmeri, P., Mendoza, C., Kallman, T.~R., \& Bautista, M.~A.\ 2002, \apjl, 577, L119
\bibitem[piconcelli (2007)]{piconcelli07} Piconcelli, E., Bianchi, S., Guainazzi, M., Fiore, F., \& Chiaberge, M.\ 2007, \aap, 466, 855
\bibitem[pounds (2003)]{pounds03} Pounds, K.~A., Reeves, 
J.~N., King, A.~R., Page, K.~L., O'Brien, P.~T., 
\& Turner, M.~J.~L.\ 2003, \mnras, 345, 705
\bibitem[pounds (2006)]{pounds06} Pounds, K.~A., \& Page, K.~L.\ 2006, \mnras, 372, 1275
\bibitem[Pounds \& Reeves(2007)]{2007MNRAS.374..823P} Pounds, K.~A., \& Reeves, J.~N.\ 2007, \mnras, 374, 823
\bibitem[Proga \& Kallman(2004)]{2004ApJ...616..688P} Proga, D., \& Kallman, T.~R.\ 2004, \apj, 616, 688
\bibitem{2005ApJ...630L.129R} Risaliti, G., Bianchi, S., Matt, G., Baldi, A., Elvis, M., Fabbiano, G., \& Zezas, A.\ 2005, ApJ, 630, L129
\bibitem[reeves (2004)]{reeves04} Reeves, J.~N., Nandra, K., George, I.~M., Pounds, K.~A., Turner, T.~J., \& Yaqoob, T.\ 2004, \apj, 602, 648
\bibitem[Reeves et al.(2009)]{2009ApJ...701..493R} Reeves, J.~N., et al.\ 2009, \apj, 701, 493
\bibitem[Rybicki \& Lightman(1979)]{1979rpa..book.....R} Rybicki, G.~B., \& Lightman, A.~P.\ 1979, Radiative Processes in Astrophysics (New York: Wiley) 
\bibitem[Sim et al.(2008)]{2008MNRAS.388..611S} Sim, S.~A., Long, K.~S., Miller, L., \& Turner, T.~J.\ 2008, \mnras, 388, 611
\bibitem[Sim et al.(2010)]{2010MNRAS.tmp..354S} Sim, S.~A., Miller, L., Long, K.~S., Turner, T.~J., \& Reeves, J.~N.\ 2010, \mnras, 354
\bibitem[Tarter et al.(1969)]{1969ApJ...156..943T} Tarter, C.~B., Tucker, W.~H., \& Salpeter, E.~E.\ 1969, \apj, 156, 943
\bibitem[Tombesi et al.(2010)]{2010A&A...521A..57T} Tombesi, F., Cappi, M., Reeves, J.~N., Palumbo, G.~G.~C., Yaqoob, T., Braito, V., \& Dadina, M.\ 2010, \aap, 521, A57; (Tombesi et al.~2010a; TA)
\bibitem[Tombesi et al.(2010)]{2010ApJ...719..700T} Tombesi, F., Sambruna, R.~M., Reeves, J.~N., Braito, V., Ballo, L., Gofford, J., Cappi, M., \& Mushotzky, R.~F.\ 2010, \apj, 719, 700; (Tombesi et al.~2010b; TB)
\bibitem[turner (2007)]{turner07} Turner, T.~J., Miller, L., Reeves, J.~N., \& Kraemer, S.~B.\ 2007, \aap, 475, 121
\bibitem[turner (2008)]{turner08} Turner, T.~J., Reeves, J.~N., Kraemer, S.~B., \& Miller, L.\ 2008, \aap, 483, 161
\bibitem{1995PASP..107..803U} Urry, C.~M., \& Padovani, P.\ 1995, PASP, 107, 803
\bibitem[Verner et al.(1996)]{1996ADNDT..64....1V} Verner, D.~A., Verner, E.~M., \& Ferland, G.~J.\ 1996, Atomic Data and Nuclear Data Tables, 64, 1

\end{thebibliography}
\end{document}